\documentclass[lettersize,journal]{IEEEtran}
\usepackage[colorlinks,
            linkcolor=blue,       %%修改此处为你想要的颜色
            anchorcolor=blue,  %%修改此处为你想要的颜色
            citecolor=blue,        %%修改此处为你想要的颜色，例如修改blue为red
            ]{hyperref}
\usepackage{cite}
\usepackage{amsmath,amssymb,amsfonts}
\usepackage{algorithmic}
\usepackage{graphicx}
\usepackage{textcomp}
\usepackage{xcolor}
\usepackage{array}
\usepackage{subfloat}
\usepackage{subfigure}

\newtheorem{definition}{Definition}

\def\BibTeX{{\rm B\kern-.05em{\sc i\kern-.025em b}\kern-.08em
    T\kern-.1667em\lower.7ex\hbox{E}\kern-.125emX}}
    
\begin{document}
\title{Age-Energy Analysis in Multi-Source Systems with Wake-up Control and Packet Management}
\author{Jie Gong,~\IEEEmembership{Member,~IEEE}, and Jiajie Huang
\thanks{J. Gong and J. Huang are with Guangdong Key Laboratory of Information Security Technology, School of Computer Science and Engineering, Sun Yat-sen University, Guangzhou 510006, China (e-mails: gongj26@mail.sysu.edu.cn; huangjj7@mail2.sysu.edu.cn).}}%
\maketitle

\begin{abstract}%
In recent years, there has been an increasing focus on real-time mobile applications, such as news updates and weather forecast. In these applications, data freshness is of significant importance, which can be measured by age-of-synchronization (AoS). At the same time, the reduction of carbon emission is increasingly required by the communication operators. Thus, how to reduce energy consumption while keeping the data fresh becomes a matter of concern. In this paper, we study the age-energy trade-off in a multi-source single-server system, where the server can turn to sleep mode to save energy. We adopt the stochastic hybrid system (SHS) method to analyze the average AoS and power consumption with three wake-up policies including N-policy, single-sleep policy and multi-sleep policy, and three packet preemption strategies, including Last-Come-First-Serve with preemption-in-Service (LCFS-S), LCFS with preemption-only-in-Waiting (LCFS-W), and LCFS with preemption-and-Queueing (LCFS-Q). The trade-off performance is analyzed via both closed-form expressions and numerical simulations. It is found that N-policy attains the best trade-off performance among all three sleep policies. Among packet management strategies, LCFS-S is suitable for scenarios with high requirements on energy saving and small arrival rate difference between sources. LCFS-Q is suitable for scenarios with high requirements on information freshness and large arrival rate difference between sources.
\end{abstract}
%p

\section{Introduction}\label{0}
\IEEEPARstart{W}{ith} the proliferation of mobile devices and the growing demand for real-time information, mobile applications have become an integral part of our daily lives. The real-time applications, such as news apps, social media platforms, and weather apps, is expected to deliver timely updates to keep users informed. The delivery of up-to-the-minute information is critical for providing a seamless user experience. On the other hand, the development of communication technology has also led to the increase of the operators' expense on electric power and hence, the increase of carbon emission. Consequently, energy consumption has become a key issue of information and communication technology. Thus, in the pursuit of improving the freshness of information, how to reduce energy consumption has also aroused widespread concern. However, there is a mutually restrictive relationship between keeping data fresh and reducing energy consumption. Intuitively, achieving high data freshness necessitates frequent updates, which, in turn, results in high energy consumption. Conversely, to conserve energy, devices need to reduce their operation time or enter low-power modes, thereby jeopardizing data freshness. Therefore, there exists a fundamental trade-off between age and energy.

To measure the freshness of the data, a new metric called age-of-information (AoI) is proposed in \cite{AOI}, which is defined as the time elapsed since the latest received update was generated. Different from the traditional delay metric, AoI captures the effects of transmission delay and data update frequency at the same time. It has been extensively studied in the literature \cite{coding1,coding3,physical1,physical2,network1,network2,network3,6,7,10,12,13}. However, AoI only indicates the age of the data in hand, but can not reveal whether the status of the source changes or not. To cope with it, Age-of-synchronization (AoS) is proposed in \cite{AOS} to measure whether the data is synchronized. It is defined as the time elapsed since the receiver's latest information became asynchronous. Unlike AoI metric, AoS is more suitable for scenarios where data updates are less frequent, because the data may be mistakenly considered to be very stale under AoI metrics even if the status of the source does not change \cite{AOS1, AOS2, AOS3}. Due to the differences between AoI and AoS, performance studies conducted on AoI may not be optimally suited for AoS. Therefore, it is of importance to study how to improve the AoS performance.

%\begin{figure}[htb]
%\centering
%\includegraphics[width=80mm]{new_system_model-eps-converted-to.pdf}
%\caption{Status update system model.}
%\label{new system model}
%\end{figure}

In this paper, we analyze AoS and power consumption of a multi-source status update system, composed of $M$ sources and a single server. Each source generates status updates according to a certain distribution. The updates are then delivered by the server with a random delivery time. Once the transmission is complete, the data at the receiver side is refreshed and synchronized to the source. When there are multiple updates in the server, preemption strategies should be considered to discard outdated updates. When the update frequency is low, there is opportunity for the server to sleep to save energy, and wake-up policies should be designed to restart the server. The main purpose of this paper is to study the age-energy trade-off of the system under different preemption strategies and wake-up policies.

\subsection{Related Work}
%As the most typical metric to describe the freshness, AoI has been extensively studied in the literature \cite{coding1,coding3,physical1,physical2,network1,network2,network3,6,7,10,12,13}. Average AoI minimization in coding was studied in \cite{coding1} and \cite{coding3}. The AoI optimization in physical layer design was studied in \cite{physical1} and \cite{physical2}. Refs.~\cite{network1, network2, network3} considered the average AoI optimization through network design. AoI performance under different queuing models was analyzed in detail, such as M/M/1, M/D/1, D/M/1 \cite{AOI}, D/G/1 \cite{6}, G/G/1/1 \cite{7}, and further studies were extended to multi-server \cite{10}, multi-data source \cite{12} and multi-hop \cite{13} scenarios. Recently, the AoS was proposed to measure whether and when the status of the source changes \cite{AOS}. It is suitable for the applications such as databases, web crawling systems and news subscription problems.
%A wireless network in which a base station sends random updates to multiple users was considered in \cite{AOS1} and \cite{AOS2}. The lower bound of the average AoS was obtained by convex optimization and random analysis, and a low complexity index-based scheduling algorithm was proposed to approach the lower bound. The scheduling of base stations broadcasting status updates to multiple nodes through shared channels was considered in \cite{AOS3}. A maximum weight strategy was proposed to minimize the weighted sum of the average AoS of all nodes.

In recent years, there are many studies on the combination of information freshness and energy consumption. The source of energy harvesting was considered in \cite{energy1}. Age optimal strategy under infinite battery \cite{energy2}, unit battery \cite{energy3} and finite battery \cite{energy4} were studied, respectively. The average AoI performance in the case of status update and random arrival of energy units was analyzed in \cite{energy6} and \cite{energy7}. The trade-off analysis between energy consumption and AoI attracted more and more attention recently \cite{energy11}, \cite{energy12}. The age-energy trade-off was revealed in error-prone channels \cite{energy8} and in fading channels \cite{energy9}, respectively. The age-energy trade-off generated by random updates without feedback was analyzed in \cite{energy10}. However, to our best knowledge, the trade-off between AoS and energy consumption is an open problem and has not been studied yet.

To analyze the data freshness over networks, the {stochastic hybrid system (SHS)} method was proposed in \cite{SHS for AoI} due to its effectiveness and simplicity. This method has been applied to various scenarios and queuing models. The average AoI of each node in a single-source multi-hop status update system was considered in \cite{SHS1}. The authors of \cite{SHS2} considered a multi-source \textit{First-Come-First-Serve (FCFS)} M/M/1 queuing model with infinite queue length. In \cite{SHS3}, a multi-server and multi-source \textit{Last-Come-First-Serve (LCFS)} queuing model was considered, and preemption in service was adopted. The authors of \cite{SHS4} considered a status update system with two sources and proposed three packet management strategies. The average AoI under each strategy was deduced by SHS. However, none of these works take into account energy consumption. In addition, applying { the SHS method} to analyze AoS also remains open. Our work tries to use SHS method to analyze energy consumption and AoS simultaneously.

\subsection{Main Results}
In this paper, {we focus on the age-energy performance analysis and comparison of some representative policies, aiming to gain some insights in theory and provide guidance for practical implementation.} We consider three packet preemption strategies, namely \textit{LCFS with preemption-in-Service (LCFS-S)}, \textit{LCFS with preemption-only-in-Waiting (LCFS-W)}, and \textit{LCFS with preemption-and-Queueing (LCFS-Q)}. In addition, we introduce the sleep model to save energy when the server is idle, and propose three wake-up policies, which are \textit{N-policy}, \textit{Single-sleep policy} and \textit{Multi-sleep policy}. Both non-ideal and ideal sleep models are considered, depending on whether idle state power consumption and wake-up time are non-zero or not. We adopt the SHS method to analyze the average AoS and power consumption under different strategies. {Distinguished from the existing works, the novelty of this paper lies in the analysis of AoS-energy tradeoff with sleep mode, adopting SHS method to analyze AoS performance, and explicit analysis and evaluation for different sleep policies and packet management strategies.} The main contributions are summarized as follows.

\begin{itemize}
\item[$\bullet$] First, we study the special case with a single source. In this case, we obtain the explicit expressions of the average AoS and the average power consumption under different wake-up policies for both non-ideal sleep model and ideal sleep model. Numerical results show that N-policy achieves the lowest AoS with the same energy consumption and vice versa, especially when the data arrival rate is high.
\item[$\bullet$] Secondly, we study the case that the number of sources is two. We obtain the explicit expressions of the average AoS and the average power consumption under different wake-up policies with LCFS-S preemption strategy. Numerical results show that LCFS-S is more energy efficient than LCFS-W (LCFS-Q). However, when the arrival rate between the sources varies greatly, the total average AoS with LCFS-S becomes large.
\item[$\bullet$] Then, when the number of sources is more than two, the explicit expressions of the average AoS and average power consumption of different preemption strategies are given for the ideal sleep model and 1-policy. Numerical results show that LFCS-S is able to attain the minimum power consumption, LCFS-Q can attain the minimum AoS, and LCFS-W attains a medium performance between AoS and power consumption.
\item[$\bullet$] Finally, we extend the service time and wake-up time to arbitrary distributions. The particular case with constant wake-up time is analyzed to illustrate that analyzing general scenarios with the SHS method is feasible.
\end{itemize}

The rest of this paper is organized as follows. The system model is shown, and the wake-up policies and preemption strategies are presented in Section \ref{II}. Section \ref{III} briefly introduces the SHS method and how to adopt SHS to analyze AoS and power consumption. The average AoS and the average power consumption for different situations are derived in Section \ref{IV}. Numerical results are presented in Section \ref{V}. Finally, Section \ref{VI} concludes the paper.

\section{System Model}\label{II}%
\begin{figure}[htb]
\centering
\includegraphics[width=60mm]{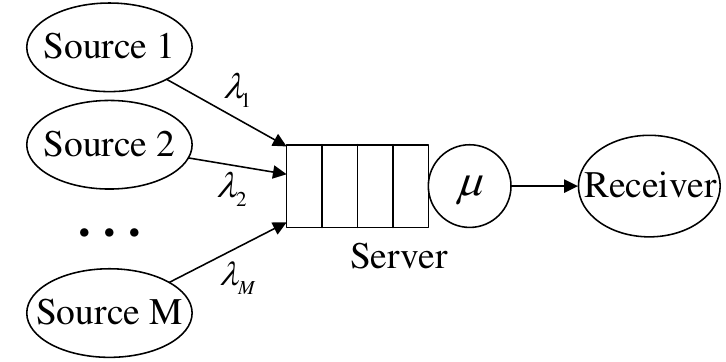}
\caption{Multi-source single-server queuing model.}
\label{system model}
\end{figure}

Consider a status update system composed of $M$ sources, a server and a receiver as shown in Fig.~\ref{system model}. Each source observes a physical process and generates status update packets to be processed by the server. It is assumed that source $i$ generates status updates following a Poisson process with parameter $\lambda_i$, and is independent with other sources. Thus, the total arrival rate of the system follows a Poisson process with  parameter $\lambda=\lambda_1+\lambda_2+\cdots+\lambda_M$. The server processes the data packets for an exponentially distributed service time with a parameter $\mu$, and then sends the completed data to a receiver. The server is energy-constrained. Hence, it looks for sleep opportunities when there are no updates from the sources.

\begin{figure}[htb]
\centering
\includegraphics[width=50mm]{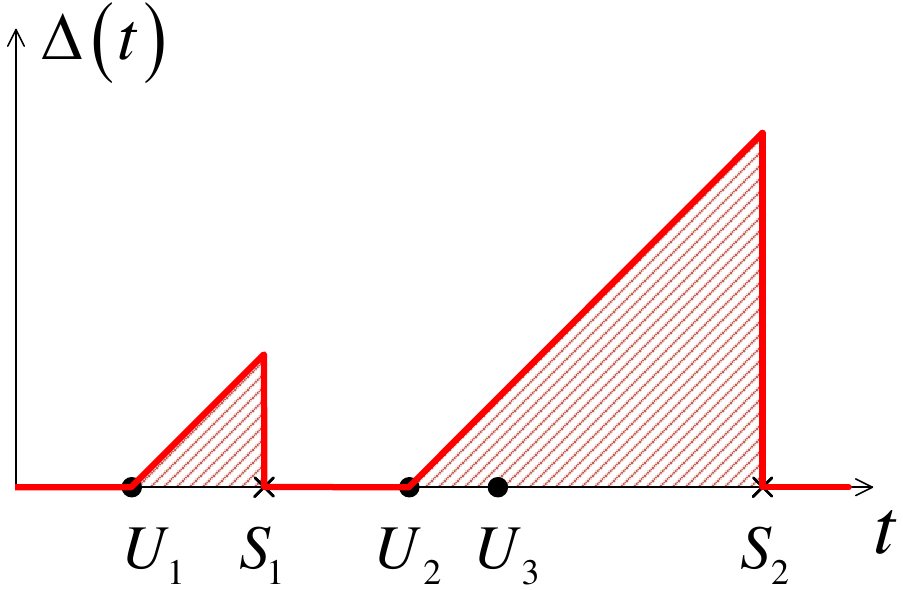}
\caption{Sample AoS path. $ \bullet $ indicates the status update at the source and $ \times $ indicates the status update at the receiver is synchronized with the source.}
\label{AoS path}
\end{figure}

We adopt the AoS \cite{AOS} as the performance metric for information freshness. Fig. \ref{AoS path} describes a sample path of AoS. Define ${U_1},{U_2}, \ldots {U_k}$ as the sequence of source update times, ${S_1},{S_2}, \ldots {S_k}$ as the sequence of times when receiver synchronizes with the source. In addition, we denote $N(t)$ as the number of synchronizations before $t$. The definition of AoS is as follows.
\begin{definition}
	\cite{AOS} \emph{Let ${u}\left( t \right)$ denote the earliest time that the source gets a status update since the last synchronization of the receiver, i.e.,
		\begin{equation}
			{u}\left( t \right) = \min \left\{ {\left. {{U_k}} \right|{U_k} > {S_{N\left( t \right)}}} \right\}.
			\label{eq01}
		\end{equation}
		The AoS at time $t$ is defined as
		\begin{equation}
			\Delta(t) = \max (t - u(t),0).
			\label{eq1}
		\end{equation}}
%		Note that if the receiver's update is the same as the source, then $\Delta(t) = 0$.}
\end{definition}

It can be seen from Fig. \ref{AoS path} that $u(S_1) = U_1$ since $U_1$ is the only update time before $S_1$. Similarly, $u(S_2) = U_2$ since $U_2$ is the first update time after the previous sync time $S_2$. The value of AoS stays zero until asynchronous between the receiver and the remote source at times $U_1$ and $U_2$. Since these points, it begins to rise steadily until it attains synchronization at $S_1$ and $S_2$. 

\begin{figure}[htb]
\centering
\includegraphics[width=60mm]{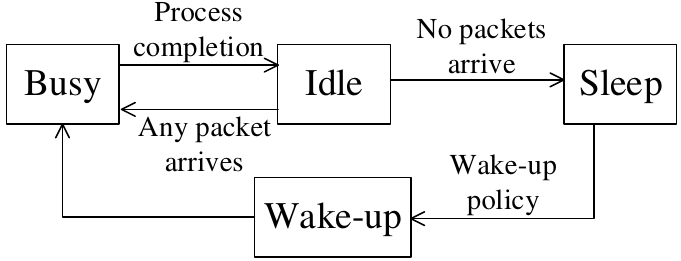}
\caption{The sleep and wake-up model of server.}
\label{sleep model}
\end{figure}

The server sleep and wake-up model is shown in Fig. \ref{sleep model}. In the \emph{busy} state, the server processes the status packets from the sources. When all the packets in the server are completely processed, the server enters the \emph{idle} state. In the idle state, the server operates at a low power level. If a new packet arrives, the server immediately enters busy state and starts working without any delay or energy cost. If no new packets arrive during the entire idle state, the server enters the \emph{sleep} state with lower power level. The duration of the idle state is assumed to follow an exponential distribution with mean $d$. Then, according to some wake-up policy, the server transfers from the sleep state to the \emph{wake-up} state to restart the device. After that, it turns to the busy state and begins to process the packet. The time cost of wake-up state is assumed to follow an exponential distribution with mean $\theta$. It is remarkable that our analysis can be extended to arbitrary distributions, which will be detailed in Sec. \ref{4D}.

The average power consumption is denoted by $\bar{P} = E[P]$. The amounts of power consumption in the four states are denoted by ${P_{\textrm{B}}}$, ${P_{\textrm{I}}}$, ${P_{\textrm{S}}}$ and ${P_{\textrm{W}}}$, respectively.

\subsection{Wake-up Policy}
The average power consumption and the average AoS depend on the wake-up policy. 
The following wake-up policies are considered.

\begin{itemize}
\item[$\bullet$] \textit{\textbf{N-policy}}: In this policy, the server immediately transfers to the wake-up state when $N$ packets arrive during the sleep state.
\item[$\bullet$] \textit{\textbf{Single-sleep policy}}: In this policy, the server turns to the wake-up state when it stays in the sleep state for a certain period of time. This period can be a constant or follows a certain distribution with mean $s$. Note that if no packets are accumulated after the wake-up process, the server keeps idle until a packet arrives.
\item[$\bullet$] \textit{\textbf{Multi-sleep policy}}: This policy can be seen as an extension of the single-sleep. In particular, the server firstly sleeps for a period of time. If no data packets arrive during this period, the server sleeps again for another period of time that follows the same distribution. Otherwise, if some packets arrive during the sleep period, the server transfers to the wake-up state.
\end{itemize}

{The motivations of considering these three policies are as follows. Firstly, the above policies are easy to implement in practice. Actually, they have been widely adopted in real systems. For instance, the wake-up radio based scheme in discontinuous reception (DRX) mode is a realization of the multi-sleep policy \cite{3GPP, rostami20:wake}. Secondly, these policies are popular \emph{vacation models} in queuing theory \cite{vacation}, which are analytically tractable. The vacation models have been extensively studied in the literature in terms of queuing performance. In this paper, we extend the analysis to AoS and energy performance. It is also shown that closed-form analytical results can be obtained.}

\subsection{Preemption Strategy}
The server deals with packets from multiple sources. To enhance the data freshness, packet preemption strategies should be deployed. Different preemption strategies change the service duration of packets from different sources, which leads to different average AoS and power consumption. We consider three preemption strategies, which are described as follows.

\begin{itemize}
\item[$\bullet$] \textit{\textbf{LCFS-S preemption strategy}}: In this strategy, any newly generated packet from any source directly preempts the packet being processed. It is equivalent to a queuing model without waiting buffer at the server. Any newly arrived packet in busy state is processed directly while the packet in service is discarded.
\item[$\bullet$] \textit{\textbf{LCFS-W preemption strategy}}: In this strategy, when a packet is being processed in the server, it can only be preempted by the packets from the same source, whereas the packets from other sources preempt the one in the queuing buffer and wait for service. It is equivalent to a queuing system with unit buffer size, and the packets from different sources need to wait after they arrive.
\item[$\bullet$] \textit{\textbf{LCFS-Q preemption strategy}}: This strategy is similar to the LCFS-W, except that the preemption in the queuing buffer only happens among packets from the same source. {It is equivalent to a FCFS preemption queue with buffer size $M-1$ for all sources. In particular, if a packet from a certain source arrives, the existing packet from the same source in the queue will be discarded, and the new one enters the end of the queue. Thus, the fresh packets from all the sources can be stored.}
\end{itemize}

In this paper, we aim to analyze the AoS and energy performance with different wake-up policies and preemption strategies. We adopt the SHS method as a unified framework for analysis, which is introduced in the next section.

\section{Analysis Framework with SHS Method}\label{III}
In this section, we first briefly introduce the SHS method and then use SHS to analyze the average AoS and the average power consumption.

\subsection{A Brief Introduction of SHS}
SHS is a kind of stochastic dynamic system which combines continuous change with discrete state variation, and the change of system structure is according to some transformation rules\cite{SHS}. In particular, the state space of the system can be divided into a discrete space and a continuous space. The evolution of the discrete state is determined by the transition or reset mapping, while the evolution of the continuous state is determined by the stochastic differential equation. The transition of discrete state is generally triggered by random events, and the probability of transition at a given time depends on the continuous and discrete components of the current SHS state. Therefore, SHS can be viewed as a piecewise deterministic Markov process in a certain sense. According to the definition of stochastic process, SHS can be expressed as
\begin{equation}
    \frac{{\textrm d{\mathbf{x}}\left( t \right)}}{{\textrm d t}} = f\left( {q\left( t \right),{\mathbf{x}}\left( t \right),t} \right) + g\left( {q\left( t \right),{\mathbf{x}}\left( t \right),t} \right)\frac{{\textrm d{\mathbf{z}}\left( t \right)}}{{\textrm d t}},
    \label{eq2}
\end{equation}%
where the $q\left( t \right) \in \mathbb{Q}$ is the discrete state, $\mathbb{Q}$ is the discrete state set,  ${\mathbf{x}}\left( t \right) \in {\mathbb{R}^{n + 1}}$ is the continuous state, ${\mathbf{z}}\left( t \right) \in {\mathbb{R}^{k}}$ describes the k-dimensional independent Brownian motion. Thus, there is a mapping $f:\mathbb{Q} \times {\mathbb{R}^{n + 1}} \times \left[ {0,\infty } \right) \to {\mathbb{R}^{n + 1}}$ and $g:\mathbb{Q} \times {\mathbb{R}^{n + 1}} \times \left[ {0,\infty } \right) \to {\mathbb{R}^{\left( {n + 1} \right) \times k}}$. Then, there is a set of transition $L$, each $l \in L$ defines a discrete transition/reset map ${\phi _l}:\mathbb{Q} \times {\mathbb{R}^{n + 1}} \times \left[ {0,\infty } \right) \to \mathbb{Q} \times {\mathbb{R}^{\left( {n + 1} \right) \times k}}$. Therefore, the state transition is 
\begin{subequations}\label{eq3}
\begin{equation}
    \left( {q'\left( t \right),{\mathbf{x'}}\left( t \right)} \right) = {\phi _l}\left( {q\left( t \right),{\mathbf{x}}\left( t \right),t} \right).
    \label{eq3a}
\end{equation}%
The corresponding transition intensity is
\begin{equation}
    {\lambda ^{\left( l \right)}}\left( {q\left( t \right),{\mathbf{x}}\left( t \right),t} \right),\quad{\lambda ^{\left( l \right)}}:\mathbb{Q} \times {\mathbb{R}^{n + 1}} \times \left[ {0,\infty } \right) \to \left[ {0,\infty } \right).
    \label{eq3b}
\end{equation}
\end{subequations}

When the system is in a discrete state, the continuous state evolves according to (\ref{eq2}). When the discrete state of the system changes from $q$ to $q'$, the continuous state jumps from $\mathbf{x}$ to $\mathbf{x'}$ according to (\ref{eq3a}), and the frequency of the transition is determined by (\ref{eq3b}). In practice, the transition intensity is generally the instantaneous rate at which the transition occurs.

\subsection{AoS and Power Consumption Analysis with SHS}

When using SHS to describe the AoS, the discrete state $q\left( t \right)$ represents the server occupancy, while the continuous state $\mathbf{x}\left( t \right)$ represents the deterministic constant slope ramp process. According to (\ref{eq2}) and (\ref{eq3}), the SHS model for AoS analysis is summarized as follows.

\begin{subequations}\label{eq4}
\begin{equation}
    f\left( {q\left( t \right),{\mathbf{x}}\left( t \right),t} \right) = {{\mathbf{b}}_q},
    \label{eq4a}
\end{equation}%
\begin{equation}
    g\left( {q\left( t \right),{\mathbf{x}}\left( t \right),t} \right) = 0,
    \label{eq4b}
\end{equation}%
\begin{equation}
    {\lambda ^{\left( l \right)}}\left( {q\left( t \right),{\mathbf{x}}\left( t \right),t} \right) = {\lambda ^{\left( l \right)}}{\delta _{{q_l},q(t)}},
    \label{eq4c}
\end{equation}%
\begin{equation}
    {\phi _l}\left( {q\left( t \right),{\mathbf{x}}\left( t \right),t} \right) = \left( {{{q'}_l}\left( t \right),{\mathbf{x}}\left( t \right){{\mathbf{A}}_l}} \right).
    \label{eq4d}
\end{equation}%
\end{subequations}
Next, we explain the notations in the above equations in detail. According to (\ref{eq4a}) and (\ref{eq4b}), the evolvement of the continuous state in each discrete state $q(t) = q$ is:
\begin{equation}
    \frac{{d{\mathbf{x}}\left( t \right)}}{{dt}} = {{\mathbf{b}}_q}.
    \label{eq5}
\end{equation}

In our model, the continuous state vector ${\mathbf{x}}\left( t \right)$ in SHS degrades to one dimension scalar ${x_0}\left( t \right)$. Note that the  AoS either increases with a slope of 1 or remains unchanged at value 0. To describe the change of AoS, we have ${{\mathbf{b}}_q}=b_q$, where  $b_q$ is a binary value. $b_q=1$ corresponds to the unit rate growth of ${x_0}\left( t \right)$, and $b_q=0$ corresponds to ${x_0}\left( t \right)$ keeping constant in state $q$.

The transition $l$ can be represented by a directed edge between states ${q_l}$ and ${q'_l}$ with a transition rate of ${\lambda ^{\left( l \right)}}$. The Cronecker delta function $\delta$ in (\ref{eq4c}) guarantees that the transition $l$ occurs only when the discrete state $q(t)$ equals to $q_l$. When a state transition occurs, the discrete state $q_l$ changes to the state $q'_l$, and the continuous state ${\mathbf{x}}\left( t \right)$ is transformed according to the binary transfer reset mapping matrix ${{\mathbf{A}}_l}$: ${\mathbf{x'}}\left( t \right)={\mathbf{x}}\left( t \right){{\mathbf{A}}_l}$ in (\ref{eq4d}). Similarly, the values of ${{\mathbf{A}}_l}={A}_l$ are only 0 and 1. ${A}_l=0$ means that the update is completed and the AoS is reduced to 0, and ${A}_l=1$ means that the AoS keeps increasing.

To obtain the average power consumption and average AoS, we need to calculate the stationary state probability of the Markov chain and the correlation vector between discrete state ${q}\left( t \right)$ and continuous state ${x_0}\left( t \right)$. Let ${{\mathbf{\pi}} _q}\left( t \right)$ denote the probability that the Markov chain is in state $q$, and ${{\mathbf{v}}_q}\left( t \right) = {v_{{q}}}\left( t \right)$ denotes the correlation between the discrete state and the continuous state. Therefore, we can obtain
\begin{equation}
    {{\mathbf{\pi}} _q}\left( t \right) = \Pr \left( {q\left( t \right) = q} \right) = {\rm E}\left[ {{\delta _{q,q\left( t \right)}}} \right],
    \label{eq6}
\end{equation}
\begin{equation}
    {{\mathbf{v}}_q}\left( t \right) = {v_{{q}}}\left( t \right) = {\rm E}\left[ {{x_0}\left( t \right){\delta _{q,q\left( t \right)}}} \right].
    \label{eq7}
\end{equation}

Let $L_q$ represent the set of all transitions from state $q$, and $L'_q$ denote the set of transitions passed in to state $q$. One of the basic assumptions for this kind of analysis is that Markov chain $q(t)$ is ergodic, otherwise the analysis of time average is meaningless. Under this assumption, the state probability vector ${\mathbf{\pi }}\left( t \right) = \left[ {{\pi _0}\left( t \right) \ldots {\pi _m}\left( t \right)} \right]$ always converges to the only constant vector ${\mathbf{\bar \pi }} = \left[ {{{\bar \pi }_0} \ldots {{\bar \pi }_m}} \right]$ which satisfies
\begin{subequations}\label{eq8}
\begin{equation}
    {\bar \pi _q}\sum\limits_{l \in {L_q}} {{\lambda ^{\left( l \right)}}}  = \sum\limits_{l \in {{L'}_q}} {{\lambda ^{\left( l \right)}}} {\bar \pi _{{q_l}}},\quad q \in \mathbb{Q},
    \label{eq8a}
\end{equation}
\begin{equation}
    \sum\limits_{q \in \mathbb{Q}} {{{\bar \pi }_q} = 1}.
    \label{eq8b}
\end{equation}
\end{subequations}

In addition, if the Markov chain of the discrete state is ergodic and stationary according to ${\mathbf{\bar \pi }}$, it has been shown in \cite[Theorem 4]{SHS for AoI} that there must be a non-negative solution such that
\begin{equation}
    {{\mathbf{\bar v}}_q}\sum\limits_{l \in {L_q}} {{\lambda ^{\left( l \right)}}}  = {{\mathbf{b}}_q}{\bar \pi _q} + \sum\limits_{l \in {{L'}_q}} {{\lambda ^{\left( l \right)}}} {{\mathbf{\bar v}}_{{q_l}}}{{\mathbf{A}}_l},\quad q \in \mathbb{Q}.
    \label{eq9}
\end{equation}
Then, the average AoS is given by 
\begin{equation}
    \bar\Delta  = \sum\limits_{q \in Q} {{v}_q}.
    \label{eq10}
\end{equation}
And the average power consumption can be obtained by 
\begin{equation}
    \bar{P} = \sum\limits_{q \in \mathbb{Q}} {{{\bar \pi }_q}{P_q}} ,\quad {P_q} \in \left\{ {{P_{\textrm{B}}},{P_{\textrm{I}}},{P_{\textrm{S}}},{P_{\textrm{W}}}} \right\},
    \label{eq11}
\end{equation}
where ${{P_{\textrm{B}}},{P_{\textrm{I}}},{P_{\textrm{S}}},{P_{\textrm{W}}}}$ represent the power consumption in busy state, idle state, sleep state and wake-up state, respectively.

\section{Age-Energy Analytical Results}\label{IV}
In this section, we show the analytical results under exponential distribution. Then, we generalize the analytical model to arbitrary distribution.

\subsection{Single source Analysis}
Firstly, we analyze the single-source case with $M = 1$.  Because only one source is considered, the three preemption strategies have the same effect. Thus, we only need to consider the impact of different wake-up policies.

\subsubsection{N-policy} \label{one source N-policy}
In N-policy, the state space of the Markov chain is $\mathbb{Q} =  \left\{B,ID,SL,1,2, \cdots ,N \right\} $. In particular, $B$ refers to busy state, $ID$ refers to idle state, $SL$ refers to sleep state without packet arrival, and $k \in \{1, 2, \cdots, N\}$ refers to the state that a total of $k$ packets arrive during the sleep state. Note that $q\left(t\right)=N$ equivalently represents wake-up state, as the server immediately turns to this state when the $N^{th}$ packet arrives. The continuous state degrades to a scalar $x_0\left(t\right) = \Delta \left(t\right)$, which is the AoS of the source. 

\begin{figure}[htb]
\centering
\includegraphics[width=40mm]{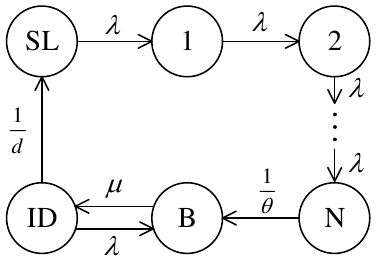}
\caption{State transition of N-policy for a single source.}
\label{Single source non-ideal N-policy}
\end{figure}

% \begin{figure*}[!t]
% \centering
% \subfigure[]{\includegraphics[width=2in]{single source non-ideal N-policy.pdf}\label{Single source non-ideal N-policy}}
% \subfigure[]{\includegraphics[width=2in]{single source non-ideal single-sleep.pdf}\label{Single source non-ideal single-sleep}}
% \subfigure[]{\includegraphics[width=2in]{single source non-ideal multi-sleep.pdf}\label{Single source non-ideal multi-sleep}}
% \caption{State transition for a single source. (a) N-policy. (b) single-sleep. (c)multi-sleep.}
% \label{one_source}
% \end{figure*}

The state transition for the discrete state $q\left( t \right)$ is shown in Fig. \ref{Single source non-ideal N-policy}. The corresponding transitions of continuous state ${\mathbf{x}}\left( t \right)$ are summarized in Table \ref{table:Single source non-ideal N-policy}, which are detailed as follows.

\begin{table}[htbp] 
\renewcommand\arraystretch{1.2}
\begin{center}
\caption{Table of Transitions for the Markov Chain in Fig. \ref{Single source non-ideal N-policy}.}  
\label{table:Single source non-ideal N-policy} 
\begin{tabular}{ | c | c | c | c | c | c | } 
\hline
     $l$   & ${{q_l} \to {{q'}_l}}$   & ${{\lambda ^{\left( l \right)}}}$   & ${{\mathbf{x}}{{\mathbf{A}}_l}}$   & ${{{\mathbf{A}}_l}}$   & ${{{{\mathbf{\bar v}}}_{{q_l}}}{{\mathbf{A}}_l}}$    \\ 
\hline
     $1$   & ${B \to ID}$   & $\mu$    & ${\left[ 0 \right]}$   & ${\left[ 0 \right]}$   & ${\left[ 0 \right]} $   \\ 
\hline
      $2$   & ${ID \to B}$   & $\lambda$    & ${\left[ {{x_0}} \right]}$   & ${\left[ 1 \right]}$   & ${\left[ {{v_{ID}}} \right]}$   \\ 
\hline
      $3$   & ${ID \to SL}$   & ${\frac{1}{d}}$   & ${\left[ {{x_0}} \right]}$   & ${\left[ 1 \right]}$   & ${\left[ {{v_{ID}}} \right]} $   \\ 
\hline
      $4$   & ${SL \to 1}$   & $\lambda$    & ${\left[ {{x_0}} \right]}$   & ${\left[ 1 \right]}$   & ${\left[ {{v_{SL}}} \right]}$   \\ 
\hline
      $5$   & ${1 \to 2}$   & $\lambda$    & ${\left[ {{x_0}} \right]}$   & ${\left[ 1 \right]}$   & ${\left[ {{v_{1}}} \right]}  $  \\ 
\hline
      $ \vdots $   & $ \vdots $   &$ \vdots $   &$ \vdots $   &$ \vdots $   &$ \vdots $        \\ 
\hline
      $N+3$   & ${N-1 \to N}$   & $\lambda$    & ${\left[ {{x_0}} \right]}$   & ${\left[ 1 \right]}$   & ${\left[ {{v_{N-1}}} \right]}$     \\ 
\hline
     $N+4$   & ${N \to B}$   & ${\frac{1}{\theta }}$   & ${\left[ {{x_0}} \right]}$   & ${\left[ 1 \right]}$   & ${\left[ {{v_{N}}} \right]}$   \\
\hline
\end{tabular}
\end{center}
\end{table}

\begin{itemize}
\item[-] \textit{$l=1$}: When a packet completes service and is delivered to the receiver, the server state changes from busy to idle. In this transition, the data on the receiver side is synchronized with the source. Thus, the AoS of the source becomes zero, i.e., $x'_0 = 0$.
\item[-] \textit{$l=2$}: When a packet arrives in the idle state, the server turns to busy state. In this case, the AoS of the source remains the same, that is, $x'_0 = x_0$, because the arrival of the packet does not reduce the AoS until it is completely processed.
\item[-] \textit{$l=3$}: If no packets arrive during the idle state, the server turns to sleep state. In this transition, $x'_0=x_0$, because no packets are processed.
\item[-] \textit{$l=4,5, \cdots, {N{+}3}$}: The state turns from state $l{-}4$ to $l{-}3$ when a new packet arrives in the sleep state, where state 0 is equivalent to state $SL$. This transition also does not change $x_0$.
\item[-] \textit{$l=N+4$}: When a total of $N$ packets arrive during sleep state, the server immediately turns to wake-up state, and then turns to busy state after a while. This transition also does not change $x_0$.
\end{itemize}

The evolution of ${\mathbf{x}}\left( t \right)$ is determined by $q\left( t \right)$. Specifically, when $q\left( t \right)=q$, we have 
\begin{equation}
    \frac{{d{\mathbf{x}}\left( t \right)}}{{dt}} = {{\mathbf{b}}_q} = \left\{ \begin{gathered}
  \left[ 1 \right],q \in \{1, 2, \cdots, N, B\}, \hfill \\
  \left[ 0 \right],q \in \{ID,SL\}. \hfill \\ 
\end{gathered}  \right.
\label{eq.bq_Single source non-ideal N-policy}
\end{equation}
The explanation for (\ref{eq.bq_Single source non-ideal N-policy}) is that $q \in \{1, 2, \cdots, N, B\}$ means that there are unprocessed packets in the system. Thus, the AoS grows at a unit rate. $q \in \{ID,SL\}$ means that there are no packets in the system. Hence, the source and the receiver are synchronized and the AoS remains at 0.

With the above conditions, we can calculate the average power consumption and average AoS by solving ${{{\bar \pi }_q}}$ and ${{\mathbf{\bar v}}_q}$. Firstly, we use (\ref{eq8a}) and (\ref{eq8b}) to calculate the stationary probability vector ${\mathbf{\bar \pi } = \left[{\bar\pi _B},{\bar\pi _{ID}},{\bar\pi _{SL}},{\bar\pi _1},{\bar\pi _2}, \cdots ,{\bar\pi _N}\right]}$. The matrix form of (\ref{eq8a}) can be expressed as ${\mathbf{\bar \pi D}} = {\mathbf{\bar \pi Q}}$, where $\mathbf{D}$ and $\mathbf{Q}$ are given as \[{\mathbf{D}} = \textrm{diag}\left[ {\mu ,\lambda  + \frac{1}{d},\underbrace {\lambda ,\lambda , \cdots ,\lambda }_{N \textrm{~elements} },\frac{1}{\theta }} \right],\] \[{\mathbf{Q}} = \left[ {\begin{array}{*{20}{c}}
  0&\mu &0&0&0&{}&0 \\ 
  \lambda &0&{\frac{1}{d}}&0&0&{}&0 \\ 
  0&0&0&\lambda &0& \vdots &0 \\ 
  0&0&0&0&\lambda &{}&0 \\ 
  {}&{}& \cdots &{}&{}& \ddots &{} \\ 
  0&0&0&0&0&{}&\lambda  \\ 
  {\frac{1}{\theta }}&0&0&0&0& \cdots &0 
\end{array}} \right].\]
With ${\mathbf{\bar \pi D}} = {\mathbf{\bar \pi Q}}$ and $\sum\nolimits_{q \in \mathbb{Q}} {{{\bar \pi }_q} = 1}$, we can obtain the stationary probability of each state as 
\begin{subequations}\label{eq:pi_Single source non-ideal N-policy}
\begin{equation}
    {{\bar \pi }_B} = A \frac{{1 + d\lambda }}{\mu },
\end{equation}
\begin{equation}
    {{\bar \pi }_{ID}} = A  d,
\end{equation}
\begin{equation}
    {{\bar \pi }_{SL}} = {{\bar \pi }_1} =  \cdots  = {{\bar \pi }_{N - 1}} = A \frac{1}{\lambda },
\end{equation}
\begin{equation}
    {{\bar \pi }_N} = A \theta,
\end{equation}
\end{subequations}
where $A = \frac{1}{{\frac{N}{\lambda } + \frac{1}{\mu } + \theta  + d\left( {1 + \frac{\lambda }{\mu }} \right)}}$.

By substituting (\ref{eq:pi_Single source non-ideal N-policy}) into (\ref{eq9}), the value of ${\mathbf{\bar v}}_q$ is obtained. Further substituting ${\mathbf{\bar v}}_q$ into (\ref{eq10}), the average AoS is given as 
\begin{equation}
    \bar \Delta  = \frac{{\frac{{1 + d\lambda }}{\mu } + \frac{\theta }{\mu } + {\theta ^2} + \frac{{N\left( {N - 1} \right)}}{{2{\lambda ^2}}} + \frac{{N - 1}}{\lambda }\left( {\theta  + \frac{1}{\mu }} \right)}}{{\frac{N}{\lambda } + \frac{1}{\mu } + \theta  + d\left( {1 + \frac{\lambda }{\mu }} \right)}}.
\end{equation}
And the average power consumption can be obtained by substituting (\ref{eq:pi_Single source non-ideal N-policy}) into (\ref{eq11}), which is expressed as 
\begin{equation}
  \bar{P} = \frac{{\frac{{1 + d\lambda }}{\mu }{P_{\textrm{B}}} + d{P_{\textrm{I}}} + \frac{N}{\lambda }{P_{\textrm{S}}} + \theta {P_{\textrm{W}}}}}{{\frac{N}{\lambda } + \frac{1}{\mu } + \theta  + d\left( {1 + \frac{\lambda }{\mu }} \right)}}. 
\end{equation}

In the ideal case that the idle state and the wake-up state are ignored, the results can be simplified by setting $\theta=0$ and $d=0$, i.e., 
\begin{equation}
    {\bar \Delta _{\textrm{Ideal}}} = \frac{{\frac{{N\left( {N - 1} \right)}}{{2{\lambda ^2}}} + \frac{{N - 1 + \lambda }}{\lambda }\frac{1}{\mu }}}{{\frac{N}{\lambda } + \frac{1}{\mu }}},
\end{equation}
\begin{equation}
    \bar{P}_{\textrm{Ideal}} = \frac{{\frac{1}{\mu }{P_{\textrm{B}}} + \frac{N}{\lambda }{P_{\textrm{S}}}}}{{\frac{N}{\lambda } + \frac{1}{\mu }}}.
\end{equation}

\subsubsection{Single-sleep policy} \label{Single-sleep policy}
In single-sleep policy, the discrete state space is $\mathbb{Q} = \left\{ {SL,SL1,WK,WK1,B,ID0,ID} \right\}$. To distinguish whether the server is busy when a new packet arrives, we define two sets of states. Specifically, the states $SL$, $WK$, $ID0$, and $ID$ respectively indicate that the server is in sleep state, wake-up state, idle state after wake-up and idle state after processing without packets in the system. Note that the difference between $ID0$ and $ID$ is that the server will turn to $SL$ if no packets arrive during $ID$ state, but not in $ID0$. The states $SL1$, $WK1$, and $B$ respectively indicate that the server is in sleep state, wake-up state, and busy state with packets in the queue or in processing. The continuous state also degrades to a scalar $x_0\left(t\right) = \Delta \left(t\right)$.

\begin{figure}[htb]
\centering
\includegraphics[width=45mm]{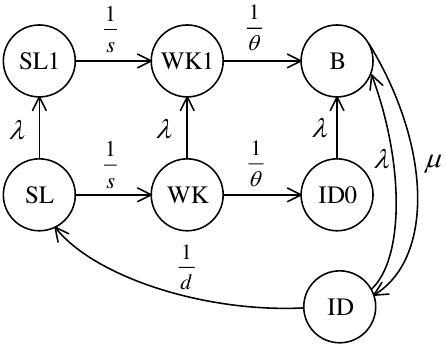}
\caption{State transition of single-sleep policy for a single source.}
\label{Single source non-ideal single-sleep}
\end{figure}

The state transition for the discrete state $q\left( t \right)$ is shown in Fig. \ref{Single source non-ideal single-sleep}. The corresponding transitions of continuous state ${\mathbf{x}}\left( t \right)$ are summarized in Table \ref{table:Single source non-ideal single-sleep}, which are detailed as follows.

\begin{table}[htbp] 
\renewcommand\arraystretch{1.2}
\begin{center}
\caption{Table of Transitions for the Markov Chain in Fig.\ref{Single source non-ideal single-sleep}.}  
\label{table:Single source non-ideal single-sleep} 
\begin{tabular}{ | c | c | c | c | c | c | } 
\hline
     $l$   & ${{q_l} \to {{q'}_l}}$   & ${{\lambda ^{\left( l \right)}}}$   & ${{\mathbf{x}}{{\mathbf{A}}_l}}$   & ${{{\mathbf{A}}_l}}$   & ${{{{\mathbf{\bar v}}}_{{q_l}}}{{\mathbf{A}}_l}}$    \\ 
\hline
     $1$   & ${SL \to SL1}$   & $\lambda$    & ${\left[ {{x_0}} \right]}$   & ${\left[ 1 \right]}$   & ${\left[ {{v_{SL}}} \right]} $   \\ 
\hline
      $2$   & ${SL \to WK}$   & ${\frac{1}{s}}$    & ${\left[ {{x_0}} \right]}$   & ${\left[ 1 \right]}$   & ${\left[ {{v_{SL}}} \right]}$   \\ 
\hline
      $3$   & ${SL1 \to WK1}$   & ${\frac{1}{s}}$   & ${\left[ {{x_0}} \right]}$   & ${\left[ 1 \right]}$   & ${\left[ {{v_{SL1}}} \right]} $   \\ 
\hline
      $4$   & ${WK \to WK1}$   & $\lambda$    & ${\left[ {{x_0}} \right]}$   & ${\left[ 1 \right]}$   & ${\left[ {{v_{WK}}} \right]}  $  \\ 
\hline
      $5$   & ${WK \to ID0}$   & ${\frac{1}{\theta }}$    & ${\left[ {{x_0}} \right]}$   & ${\left[ 1 \right]}$   & ${\left[ {{v_{WK}}} \right]}$   \\
\hline
      $6$   & ${WK1 \to B}$   & ${\frac{1}{\theta }}$    & ${\left[ {{x_0}} \right]}$   & ${\left[ 1 \right]}$   & ${\left[ {{v_{WK1}}} \right]}$   \\
\hline
      $7$   & ${ID0 \to B}$   & $\lambda$    & ${\left[ {{x_0}} \right]}$   & ${\left[ 1 \right]}$   & ${\left[ {{v_{ID0}}} \right]}$   \\
\hline
      $8$   & $ {B \to ID} $   &$ \mu $   &$ {\left[ 0 \right]} $   &$ {\left[ 0 \right]} $   &$ {\left[ 0 \right]} $        \\ 
\hline
      $9$   & ${ID \to B}$   & $\lambda$    & ${\left[ {{x_0}} \right]}$   & ${\left[ 1 \right]}$   & ${\left[ {{v_{ID}}} \right]}$     \\ 
\hline
     $10$   & ${ID \to SL}$   & ${\frac{1}{d }}$   & ${\left[ {{x_0}} \right]}$   & ${\left[ 1 \right]}$   & ${\left[ {{v_{ID}}} \right]}$   \\
\hline
\end{tabular}
\end{center}
\end{table}

\begin{itemize}
\item[-] \textit{$l=1,4$}: Packets arrive when the server is in the sleep or wake-up state. In these transition, the server turns into state SL1 or WK1.
\item[-] \textit{$l=2,3$}: After the server stays in sleep state for a period of time with mean $s$, it turns into wake-up state.
\item[-] \textit{$l=5$}: If no packets arrive during WK state, it turns into ID0 state to keep idle and wait for a packet arrival.
\item[-] \textit{$l=6$}: After the server stays in WK1 state for a period of time with mean $\theta$, it turns into busy state.
\item[-] \textit{$l=7$}: Packets arrive when the server is in ID0 state. In this transition, the server immediately turns into busy state.
\item[-] \textit{$l=8$}: When a packet completes service and is delivered to the receiver, the server state changes from busy to idle.
\item[-] \textit{$l=9,10$}: The same as $l=2,3$ in Table \ref{table:Single source non-ideal N-policy}.
\end{itemize}

Similar to the N-policy, the AoS turns to 0 only in the transition with the completion of packet processing, in which case, the data on the receiver side is synchronized with the source. In other transitions, the AoS does not change. The evolution of ${\mathbf{x}}\left( t \right)$ is determined by the discrete state $q\left( t \right)$. Specifically, when $q\left( t \right)=q$, we have 
\begin{equation}
    \frac{{d{\mathbf{x}}\left( t \right)}}{{dt}} = {{\mathbf{b}}_q} = \left\{ \begin{gathered}
  \left[ 1 \right], \quad q \in \{SL1, WK1, B\}, \hfill \\
  \left[ 0 \right], \quad q \in \{SL, WK, ID0, ID\}. \hfill \\ 
\end{gathered}  \right.
\label{eq.bq_Single source non-ideal single-policy}
\end{equation}
The explanation for (\ref{eq.bq_Single source non-ideal single-policy}) is that $q \in \{SL1, WK1, B\}$ means that there are unprocessed packets in the system. Thus, the AoS grows at a unit rate. $q \in \{SL, WK, ID0, ID\}$ means that there are no packets in the system. Hence, the AoS remains constant at 0.

With the above conditions, we can calculate the average AoS and power consumption similar to Sec. \ref{one source N-policy}. The results are given as follows.

\begin{equation}
    \bar \Delta   = \frac{{\lambda C}}{{\mu B}},
\end{equation}
\begin{equation}
  \bar{P} = \frac{{s{P_{\textrm{S}}} + \theta {P_{\textrm{W}}} + ( {d + \frac{1}{{D}}} ){P_{\textrm{I}}} + \frac{{d\lambda  + 1}}{\mu }{P_{\textrm{B}}}}}{{\frac{B}{\mu D}}},
\end{equation}
where  $
  B = \mu  + \lambda  + ds\theta {\lambda ^4} + \left( {d + s + \theta } \right)\left( {{\lambda ^2} + \mu \lambda  + s\theta \mu {\lambda ^3}} \right) + \left( {{s^2} + {\theta ^2} + 2s\theta  + ds + d\theta } \right)\mu {\lambda ^2}  + \left( {ds + s\theta  + \theta d} \right){\lambda ^3}$, \hfill \\ $ C = \left( {{\mu ^2}{\lambda ^2}s\theta  {+} \mu \lambda } \right)\left( {{s^2} {+} s\theta  {+} {\theta ^2}} \right) {+} {\mu ^2}\lambda \left( {{s^3} {+} {s^2}\theta  {+} s{\theta ^2} {+} {\theta ^3}} \right) {+}$  \\ $\mu {\lambda ^2}s\theta \left( {s {+} \theta } \right) {+} {\lambda ^3}ds\theta {+} {\lambda ^2}\left( {ds {+} s\theta  {+} \theta d} \right) {+} \lambda \left( {d {+} s {+} \theta } \right) {+} 1 $, and $ D = \lambda \left( {s\lambda  {+} 1} \right)\left( {\theta \lambda  {+} 1} \right)$.

In the ideal case with $\theta=0$ and $d=0$, we have
\begin{equation}
    {\bar \Delta _{\textrm{Ideal}}} = \frac{{{\mu ^2}{s^3}{\lambda ^2} + \mu {s^2}{\lambda ^2} + s{\lambda ^2} + \lambda }}{{\mu \left( {\mu {s^2}{\lambda ^2} + s{\lambda ^2} + \mu s\lambda  + \mu  + \lambda } \right)}},
\end{equation}
\begin{equation}
    \bar{P}_{\textrm{Ideal}} = \frac{{{\mu s\lambda \left( {s\lambda  + 1} \right)}{P_{\textrm{S}}} + \mu {P_{\textrm{I}}} + \lambda \left( {s\lambda  + 1} \right){P_{\textrm{B}}}}}{{\mu {s^2}{\lambda ^2} + s{\lambda ^2} + \mu s\lambda  + \mu  + \lambda }}.
\end{equation}

\subsubsection{Multi-sleep policy}
In multi-sleep policy, the discrete state space is $\mathbb{Q} = \left\{ {SL,SL1,WK,B,ID} \right\}$. The definitions of these states are the same as in \ref{Single-sleep policy}. The continuous state also degrades to a scalar $x_0\left(t\right) = \Delta \left(t\right)$. 

\begin{figure}[htb]
\centering
\includegraphics[width=40mm]{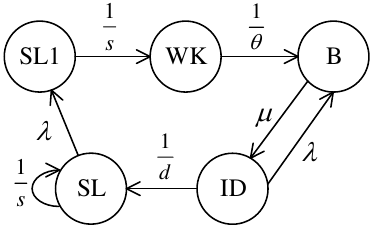}
\caption{State transition of multi-sleep policy for a single sources.}
\label{Single source non-ideal multi-sleep}
\end{figure}

The state transition for the discrete state $q\left( t \right)$ is shown in Fig. \ref{Single source non-ideal multi-sleep}. The corresponding transitions of continuous state ${\mathbf{x}}\left( t \right)$ are summarized in Table \ref{table:Single source non-ideal multi-sleep}, which is very similar to Table \ref{table:Single source non-ideal single-sleep}. $l=1,3,4,5,6,7$ is the same as $l=1,3,5,8,9,10$ in Table \ref{table:Single source non-ideal single-sleep}. The main difference is the self-transition $l = 2$, which means that the server sleeps again after one sleep period if there is no packets arrival during this period.

\begin{table}[htbp] 
\renewcommand\arraystretch{1.2}
\begin{center}
\caption{Table of Transitions for the Markov Chain in Fig. \ref{Single source non-ideal multi-sleep}.}  
\label{table:Single source non-ideal multi-sleep} 
\begin{tabular}{ | c | c | c | c | c | c | } 
\hline
     $l$   & ${{q_l} \to {{q'}_l}}$   & ${{\lambda ^{\left( l \right)}}}$   & ${{\mathbf{x}}{{\mathbf{A}}_l}}$   & ${{{\mathbf{A}}_l}}$   & ${{{{\mathbf{\bar v}}}_{{q_l}}}{{\mathbf{A}}_l}}$    \\ 
\hline
     $1$   & ${SL \to SL1}$   & $\lambda$    & ${\left[ {{x_0}} \right]}$   & ${\left[ 1 \right]}$   & ${\left[ {{v_{SL}}} \right]} $   \\ 
\hline
      $2$   & ${SL \to SL}$   & ${\frac{1}{s}}$    & ${\left[ {{x_0}} \right]}$   & ${\left[ 1 \right]}$   & ${\left[ {{v_{SL}}} \right]}$   \\ 
\hline
      $3$   & ${SL1 \to WK}$   & ${\frac{1}{s}}$   & ${\left[ {{x_0}} \right]}$   & ${\left[ 1 \right]}$   & ${\left[ {{v_{SL1}}} \right]} $   \\ 
\hline
      $4$   & ${WK \to B}$   & ${\frac{1}{\theta }}$    & ${\left[ {{x_0}} \right]}$   & ${\left[ 1 \right]}$   & ${\left[ {{v_{WK}}} \right]}$   \\
\hline
      $5$   & $ {B \to ID} $   &$ \mu $   &$ {\left[ 0 \right]} $   &$ {\left[ 0 \right]} $   &$ {\left[ 0 \right]} $        \\ 
\hline
      $6$   & ${ID \to B}$   & $\lambda$    & ${\left[ {{x_0}} \right]}$   & ${\left[ 1 \right]}$   & ${\left[ {{v_{ID}}} \right]}$     \\ 
\hline
      $7$   & ${ID \to SL}$   & ${\frac{1}{d }}$   & ${\left[ {{x_0}} \right]}$   & ${\left[ 1 \right]}$   & ${\left[ {{v_{ID}}} \right]}$   \\
\hline
\end{tabular}
\end{center}
\end{table}

Similarly, the average AoS and average power consumption can be obtained as
\begin{equation}
    \bar \Delta  = \frac{{\lambda \left( {{s^2}{\mu ^2} + s\theta {\mu ^2} + s\mu  + {\theta ^2}{\mu ^2} + \theta \mu  + d\lambda  + 1} \right)}}{{\mu \left( {\mu  + \lambda  + d{\lambda ^2} + d\mu \lambda  + s\mu \lambda  + \theta \mu \lambda } \right)}},
\end{equation}
\begin{equation}
  \bar{P} = \frac{{\mu \left( {s\lambda  + 1} \right){P_{\textrm{S}}} + \theta \mu \lambda {P_{\textrm{W}}} + \lambda \left( {d\lambda  + 1} \right){P_{\textrm{B}}} + d\mu \lambda {P_{\textrm{I}}}}}{{\mu  + \lambda  + d{\lambda ^2} + d\mu \lambda  + s\mu \lambda  + \theta \mu \lambda }}. \\
\end{equation}

The results of ideal case can be obtained by setting $\theta=0$ and $d=0$ in the above equations, i.e., 
\begin{equation}
    {\bar \Delta _{\textrm{Ideal}}} = \frac{{\lambda \left( {{s^2}{\mu ^2} + s\mu  + 1} \right)}}{{\mu \left( {\mu  + \lambda  + s\mu \lambda } \right)}},
\end{equation}
\begin{equation}
    \bar{P}_{\textrm{Ideal}} = \frac{{\mu \left( {s\lambda  + 1} \right){P_{\textrm{S}}} + \lambda {P_{\textrm{B}}}}}{{\mu  + \lambda  + s\mu \lambda }}.
\end{equation}

\subsection{Two sources Analysis}
Next, we analyze the two-sources case with $M = 2$. In this case, the preemption strategies LCFS-W and LCFS-Q become the same. However, it is difficult to obtain the closed-form expressions for them. Thus, we mainly focus on the LCFS-S preemption strategy with different wake-up policies. The analysis of LCFS-W (LCFS-Q) is presented later via numerical simulations. Since the two sources are symmetric, we only need to analyze the performance of one source. In the following, we analyze the state transition of source 1. The source 2 can be analyzed in the same way.
% The performance of LCFS-W will be shown in numerical results via simulations.

\subsubsection{N-policy} \label{N}

\begin{figure}[htb]
\centering
\includegraphics[width=65mm]{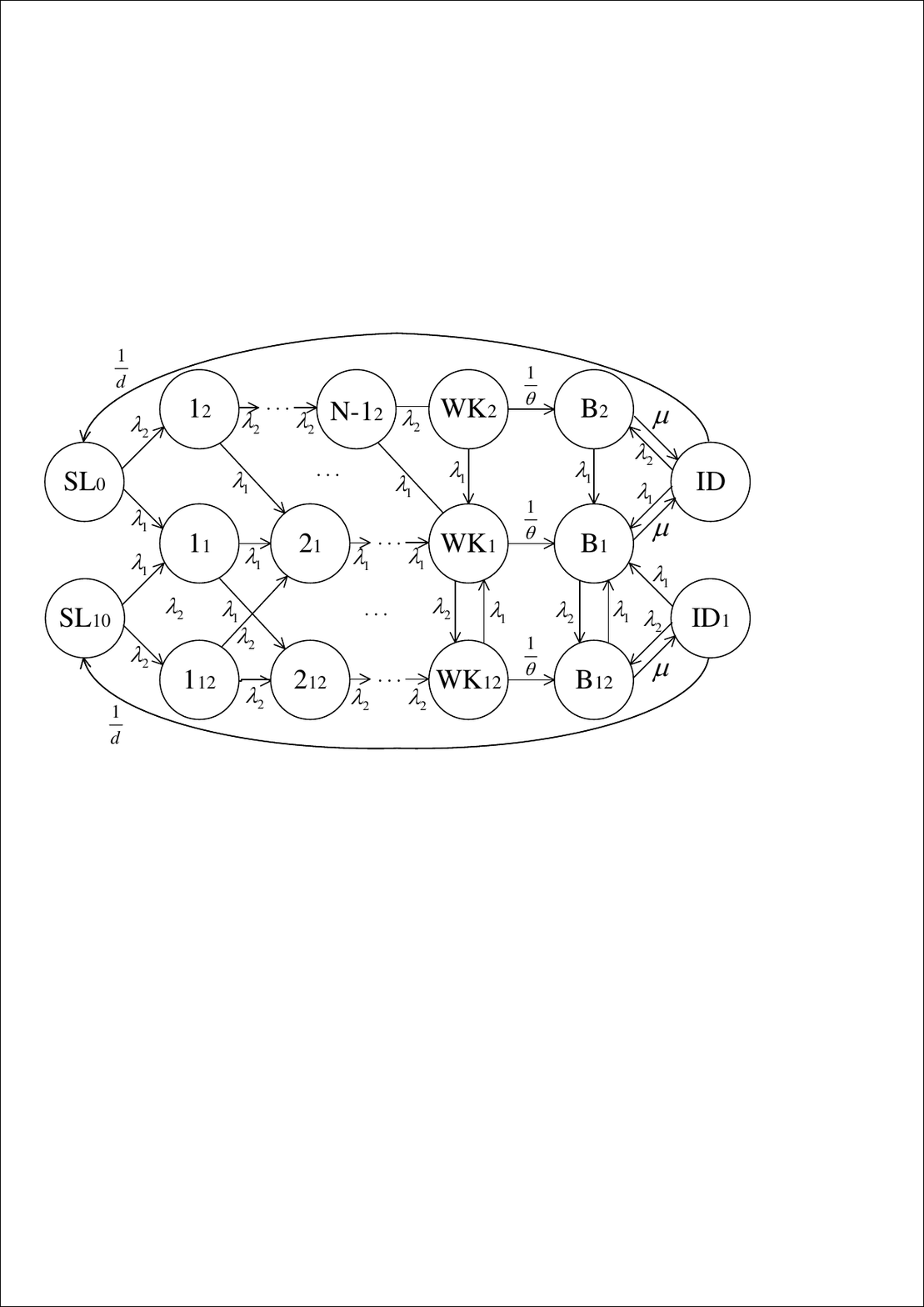}
\caption{State transition of N-policy for two sources.}
\label{two source non-ideal N-packets}
\end{figure}

% \begin{figure*}[!t]
% \centering
% \subfigure[]{\includegraphics[width=2.3in]{two source non-ideal N-policy.eps}\label{two source non-ideal N-packets}}
% \subfigure[]{\includegraphics[width=1.7in]{two source non-ideal single-sleep.pdf}\label{two source non-ideal single-sleep}}
% \subfigure[]{\includegraphics[width=2.3in]{two source non-ideal multi-sleep.eps}\label{two source non-ideal multi-sleep}}
% \caption{State transition for two sources. (a) N-policy. (b) single-sleep. (c) multi-sleep.}
% \label{two_source}
% \end{figure*}

In N-policy, the Markov chain for the discrete state $q\left( t \right)$ and the transitions between states are shown in Fig. \ref{two source non-ideal N-packets}. It can be regarded as three rings in Fig. \ref{Single source non-ideal N-policy} arranged side by side and connected to each other. The subscript "0" indicates that no packets are in the system. The subscript ``1" indicates that a packet of source 1 is in the system at this time. The subscript ``2" indicates that a packet of source 2 is in the system and source 1 is successfully synchronized. The subscript ``12" indicates that a packet of source 2 is in the system while source 1 is not synchronized as its packet is preempted in service. The subscript ``10" means that there are no packets in the system but the source 1 is not synchronized.  This happens when the packet of source 1 is preempted by that of source 2, which then finishes processing. In summary, the state with the subscript ``1" indicates that the data of the source 1 is out of synchronization. In these states, the AoS grows linearly with rate 1. The state without the subscript ``1" means that the data of the source 1 is synchronized, and the AoS remains zero. The transition among states in each ring is similar to the case of single source.

With the above SHS model, the average power consumption and average AoS can be calculated in the same way. The derived results turn to be complex. For simplicity, we present the result for the ideal sleep model when $N=1$ as follows.
\begin{equation} \label{eq:N1two}
    {\bar \Delta _{1, \textrm{Ideal}}} = \frac{{\lambda}^2+\mu\lambda_{2}}{\mu\lambda_{1}\left(\mu+\lambda\right)},
\end{equation}
\begin{equation}
    \bar{P}_{\textrm{Ideal}} = \frac{{ \mu P_S + \lambda P_B}}{{\mu  + {\lambda } }}.
\end{equation}

\subsubsection{Single-sleep policy}

\begin{figure}[htb]
\centering
\includegraphics[width=45mm]{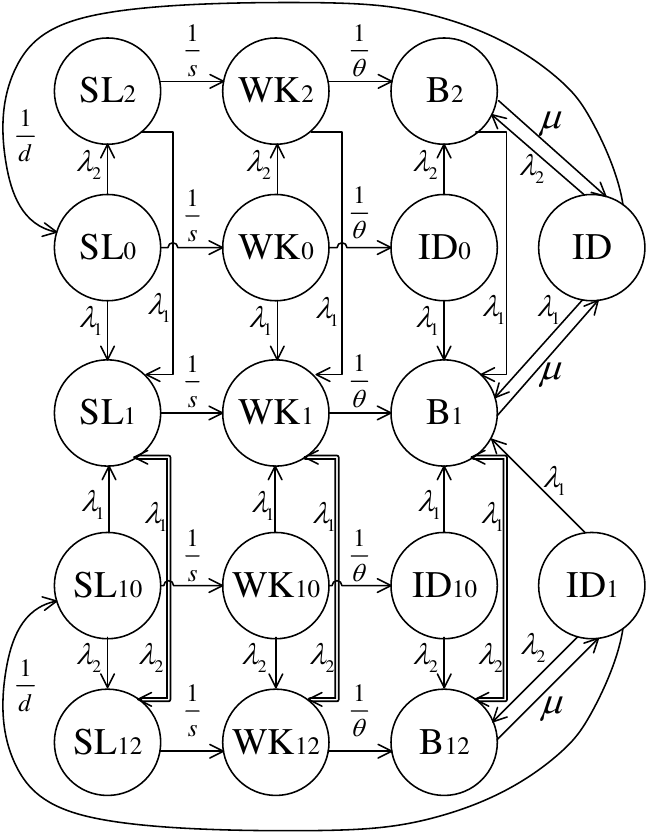}
\caption{State transition of single-sleep policy for two sources.}
\label{two source non-ideal single-sleep}
\end{figure}

In single-sleep policy, the Markov chain for the discrete state $q\left( t \right)$ and the transitions between states are shown in Fig. \ref{two source non-ideal single-sleep}. It can be regarded as an expansion of Fig. \ref{Single source non-ideal single-sleep}. The meanings of the subscripts are the same as the N-policy. The transition among states is similar to the case of single source.

As the results are over-complex, we simply present that for the ideal case as follows.
\begin{equation}\label{A1}
    {{\bar \Delta }_{1, \textrm{Ideal}}} = \frac{{\sum\limits_{i = 0}^7 {{\lambda ^i}{\eta _i}} }}{{\mu {\lambda _1}\left( {\mu  + \lambda } \right)\left( {s\lambda  + 1} \right)\sum\limits_{i = 0}^4 {{\lambda ^i}{\gamma _i}}}},
\end{equation}
\begin{equation}
    \bar{P}_{\textrm{Ideal}} = \frac{{\mu s\lambda \left( {s\lambda  + 1} \right){P_\textrm{S}} + \mu {P_\textrm{I}} + \left( {s{\lambda ^2} + \lambda } \right){P_\textrm{B}}}}{{\left( {\mu {s^2} + s} \right){\lambda ^2} + \left( {\mu s + 1} \right)\lambda  + \mu }},
\end{equation}
where the parameters are detailed in Appendix A.

\subsubsection{Multi-sleep policy}

\begin{figure}[htb]
\centering
\includegraphics[width=60mm]{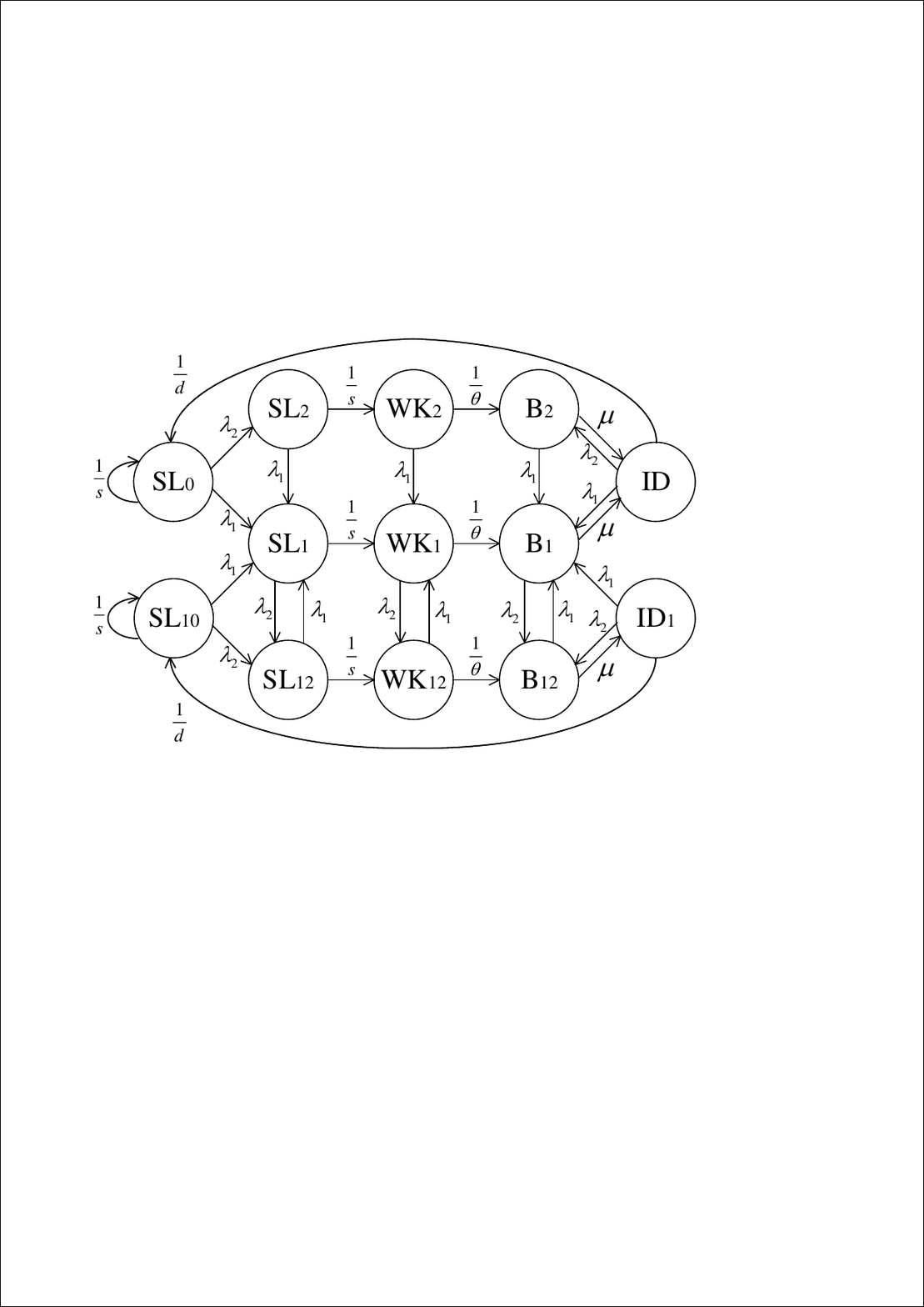}
\caption{State transition of multi-sleep policy for two sources.}
\label{two source non-ideal multi-sleep}
\end{figure}

In multi-sleep policy, the Markov chain for the discrete state $q\left( t \right)$ and the transitions between states are shown in Fig. \ref{two source non-ideal multi-sleep}. It can be regarded as three rings in Fig. \ref{Single source non-ideal multi-sleep} arranged side by side and connected to each other. The meanings of the subscripts are the same as the N-policy. The transition among states is similar to the case of single source.

As the results are over-complex, we simply present that for the ideal case as follows.
\begin{equation}\label{A2}
    {{\bar \Delta }_{1, \textrm{Ideal}}} = \frac{{\sum\limits_{i = 0}^5 {{\lambda ^i}{\eta _i}} }}{{\mu {\lambda _1}\left( {\mu  + \lambda } \right)\left( {s\lambda  + 1} \right)A\left( {s{\lambda _1}^2 + s{\lambda _1}{\lambda _2} + A} \right)}},
\end{equation}
\begin{equation}
    \bar{P}_{\textrm{Ideal}} = \frac{{\mu \left( {s\lambda  + 1} \right){P_{\textrm{S}}} + \lambda {P_{\textrm{B}}}}}{{\mu  + \lambda  + s\mu \lambda }},
\end{equation}
where the parameters are detailed in Appendix B.

\subsection{Three and More Sources}

In this subsection, we analyze the performance for three or more sources. As shown below, the analytical results become over-complex. Nevertheless, the SHS analysis framework is still applicable, either theoretically or numerically. To illustrate how to analyze the performance of more than two sources, we focus on the simple ideal sleep model. In particular, the idle state and the wake-up state are ignored. Therefore, the server can transit between busy state and sleep state without time and energy cost. Besides, for wake-up policy, we focus on the N-policy with $N=1$. Taking three sources as an example, we compare the differences among the aforementioned preemption strategies. 

\subsubsection{LCFS-S}
\begin{figure}[htb]
\centering
\includegraphics[width=35mm]{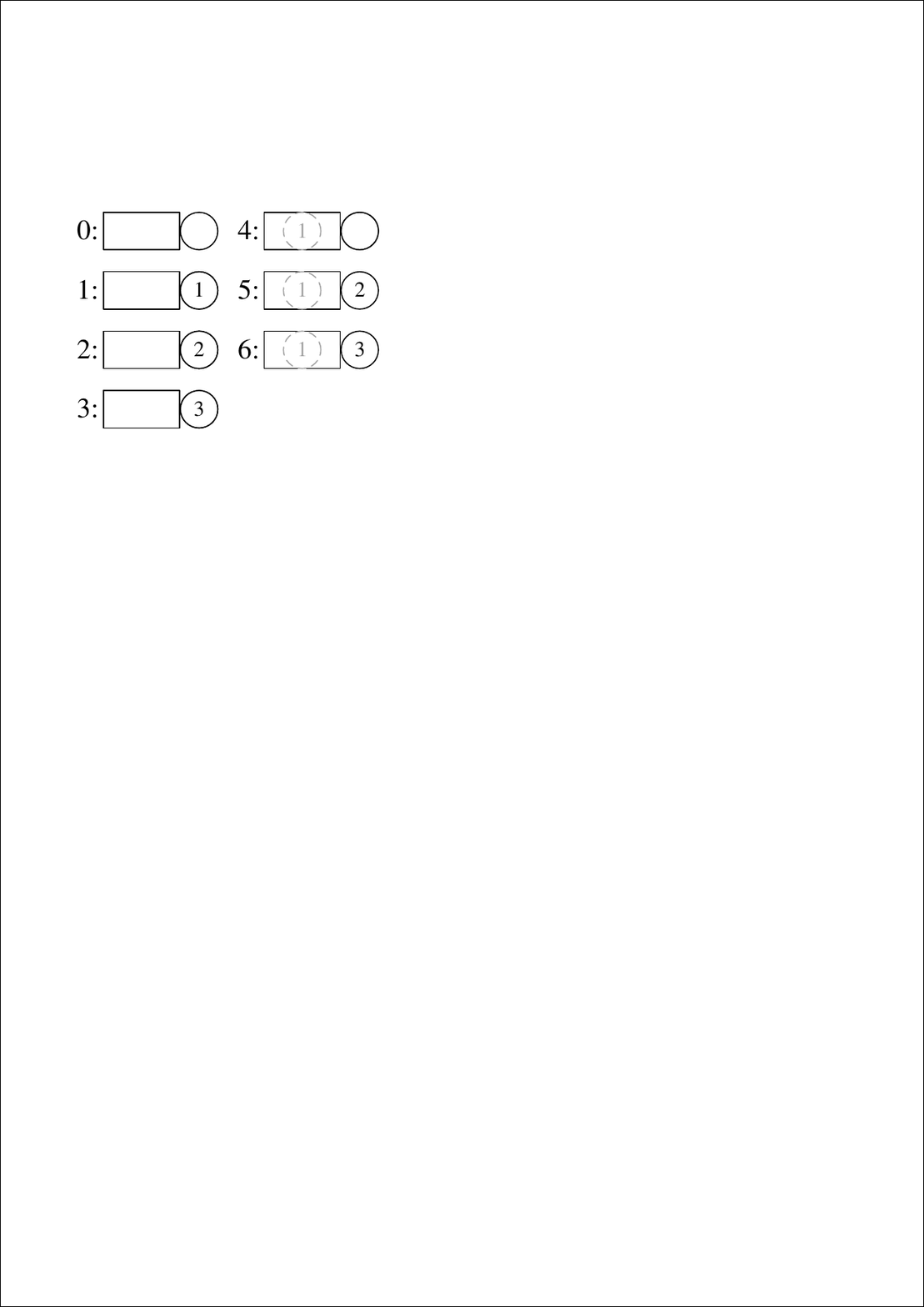}
\caption{Illustration of states with LCFS-S for 3 sources case. The rectangle on the left represents queue, the circle on the right represents the server, and the grayed circle inside the queue represents that the data from source 1 is unsynchronized.}
\label{Three source ideal LCFS-S}
\end{figure}

For the system with 1-policy, LCFS-S and ideal model, we need to define seven server states to analyze the performance of source 1, as illustrated in Fig. \ref{Three source ideal LCFS-S}. The number in the server indicates the index of source from which a packet is processing. The vacuum queue means that source 1 is synchronized. The queue with a dashed circle indicates that the data from source 1 is unsynchronized. This happens when the packet of source 1 is preempted by the packets from other sources. Thus, the server may be idle without synchronization of source 1.

% \begin{figure*}[!t]
% \centering
% \subfigure[]{\includegraphics[width=1.16in]{LCFS-S.eps}\label{Three source ideal LCFS-S}}\hspace{5.5mm}
% \subfigure[]{\includegraphics[width=2.4in]{LCFS-W.eps}\label{Three source ideal LCFS-W}}\hspace{5.5mm}
% \subfigure[]{\includegraphics[width=2.4in]{LCFS-Q.eps}\label{Three source ideal LCFS-Q}}
% \caption{Illustration of states with LCFS-Q for three sources case. The rectangle on the left represents the queue, the circle on the right represents the server, the solid circle represents that packets in queue, and the gray circle represents that the data from source 1 is unsynchronized. (a) LCFS-S. (b) LCFS-W. (c) LCFS-Q.}
% \label{three_source}
% \end{figure*}

\begin{table}[htbp] 
\renewcommand\arraystretch{1.2}
\begin{center}
\caption{Table of Transitions for the Markov Chain in Fig. \ref{Three source ideal LCFS-S}}  
\label{table:Three source ideal LCFS-S} 
\begin{tabular}{ | c | c | c | c | c | c | } 
\hline
     $l$   & ${{q_l} \to {{q'}_l}}$   & ${{\lambda ^{\left( l \right)}}}$   & ${{\mathbf{x}}{{\mathbf{A}}_l}}$   & ${{{\mathbf{A}}_l}}$   & ${{{{\mathbf{\bar v}}}_{{q_l}}}{{\mathbf{A}}_l}}$    \\ 
\hline
     $1$   & ${0 \to 1}$   & $\lambda_{1}$    & ${\left[ {{x_0}} \right]}$   & ${\left[ 1 \right]}$   & ${\left[ {{v_{0}}} \right]} $   \\ 
\hline
      $2$   & ${0 \to 2}$   & $\lambda_{2}$    & ${\left[ {{x_0}} \right]}$   & ${\left[ 1 \right]}$   & ${\left[ {{v_{0}}} \right]}$   \\ 
\hline
      $3$   & ${0 \to 3}$   & $\lambda_{3}$    & ${\left[ {{x_0}} \right]}$   & ${\left[ 1 \right]}$   & ${\left[ {{v_{0}}} \right]}$   \\ 
\hline
      $4$   & ${1 \to 0}$   & $\mu$    & ${\left[ 0 \right]}$   & ${\left[ 0 \right]}$   & ${\left[ {{0}} \right]}  $  \\ 
\hline
      $5$   & ${1 \to 5}$   & $\lambda_{2}$    & ${\left[ {{x_0}} \right]}$   & ${\left[ 1 \right]}$   & ${\left[ {{v_{1}}} \right]}$   \\
\hline
      $6$   & ${1 \to 6}$   & $\lambda_{3}$      & ${\left[ {{x_0}} \right]}$   & ${\left[ 1 \right]}$   & ${\left[ {{v_{1}}} \right]}$   \\
\hline
      $7$   & ${2 \to 0}$   & $\mu$    & ${\left[ {{x_0}} \right]}$   & ${\left[ 1 \right]}$   & ${\left[ {{v_{2}}} \right]}$   \\
\hline
      $8$   & $ {2 \to 1} $   & $\lambda_{1}$  & ${\left[ {{x_0}} \right]}$   & ${\left[ 1 \right]}$   & ${\left[ {{v_{2}}} \right]}$   \\
\hline
      $9$   & ${2 \to 3}$   & $\lambda_{3}$    & ${\left[ {{x_0}} \right]}$   & ${\left[ 1 \right]}$   & ${\left[ {{v_{2}}} \right]}$     \\ 
\hline
     $10$   & ${3 \to 0}$   & $\mu$    & ${\left[ {{x_0}} \right]}$   & ${\left[ 1 \right]}$   & ${\left[ {{v_{3}}} \right]}$   \\
\hline
     $11$   & ${3 \to 1}$   & $\lambda_{1}$    & ${\left[ {{x_0}} \right]}$   & ${\left[ 1 \right]}$   & ${\left[ {{v_{3}}} \right]} $   \\ 

\hline
      $12$   & ${3 \to 2}$   & $\lambda_{2}$    & ${\left[ {{x_0}} \right]}$   & ${\left[ 1 \right]}$   & ${\left[ {{v_{3}}} \right]}$   \\ 
\hline
      $13$   & ${4 \to 1}$   & $\lambda_{1}$    & ${\left[ {{x_0}} \right]}$   & ${\left[ 1 \right]}$   & ${\left[ {{v_{4}}} \right]}$   \\ 
\hline
      $14$   & ${4 \to 5}$   & $\lambda_{2}$    & ${\left[ {{x_0}} \right]}$   & ${\left[ 1 \right]}$   & ${\left[ {{v_{4}}} \right]}$   \\ 
\hline
      $15$   & ${4 \to 6}$   & $\lambda_{3}$   & ${\left[ {{x_0}} \right]}$   & ${\left[ 1 \right]}$   & ${\left[ {{v_{4}}} \right]}$   \\
\hline
      $16$   & ${5 \to 1}$   & $\lambda_{1}$      & ${\left[ {{x_0}} \right]}$   & ${\left[ 1 \right]}$   & ${\left[ {{v_{5}}} \right]}$   \\ 
\hline
      $17$   & ${5 \to 4}$   & $\mu$   & ${\left[ {{x_0}} \right]}$   & ${\left[ 1 \right]}$   & ${\left[ {{v_{5}}} \right]}$   \\
\hline
      $18$   & $ {5 \to 6} $   & $\lambda_{3}$   &  ${\left[ {{x_0}} \right]}$   & ${\left[ 1 \right]}$   & ${\left[ {{v_{5}}} \right]}$   \\
\hline
      $19$   & ${6 \to 1}$   & $\lambda_{1}$    & ${\left[ {{x_0}} \right]}$   & ${\left[ 1 \right]}$   & ${\left[ {{v_{6}}} \right]}$     \\ 
\hline
     $20$   & ${6 \to 4}$   & $\mu$    & ${\left[ {{x_0}} \right]}$   & ${\left[ 1 \right]}$   & ${\left[ {{v_{6}}} \right]}$   \\
\hline
     $21$   & ${6 \to 5}$   & $\lambda_{2}$   & ${\left[ {{x_0}} \right]}$   & ${\left[ 1 \right]}$   & ${\left[ {{v_{6}}} \right]} $   \\ 
\hline
\end{tabular}
\end{center}
\end{table}

The evolution of ${\mathbf{x}}\left( t \right)$ depends on whether the system contains the packet of source 1, i.e., $b_q=0$ for $q\in\{0,2,3\}$ and $b_q = 1$ for the rest. The transitions between states are shown in Table \ref{table:Three source ideal LCFS-S}. The transitions with rate $\lambda_{1}$, $\lambda_{2}$, $\lambda_{3}$ represent that a packet from source 1, source 2, source 3 arrives. Because no packets finish processing in these transitions, we have $x'_0=x_0$. The transitions with rate $\mu$ represent that the server completes process and delivers the packet to the receiver. Among these transitions, only $l=4$ makes $x'_0=0$, because the data from source 1 is synchronized. 

With the above analysis, we can calculate the average power consumption and average AoS as follows.
\begin{equation} \label{eq:N1three}
    {\bar \Delta_{1, \textrm{Ideal}} } = \frac{{\lambda}^2+\mu\lambda_{2}+\mu\lambda_{3}}{\mu\lambda_{1}\left(\mu+\lambda\right)},
\end{equation}
\begin{equation}
    \bar{P}_{\textrm{Ideal}} = \frac{{\mu {P_\textrm{S}} +  {{\lambda }} {P_\textrm{B}}}}{{\mu  + {\lambda}}}.
\end{equation}

{In fact, the results of LCFS-S for more than two sources can be recursively obtained from those for two-source case. The reason is that, as the LCFS-S preemption does not distinguish sources, one can view source 2 to $M$ as a single ``super source" with arrival rate $\lambda_2 + \cdots + \lambda_M$ based on the additive property of Poisson process. Then, the performance of source 1 in the multi-source system is the same as that in the equivalent two-source system composed of source 1 and a super source. It can be verified by comparing \eqref{eq:N1two} and \eqref{eq:N1three}, where $\lambda_2$ in \eqref{eq:N1two} for two-source case is replaced by $\lambda_2 + \lambda_3$ for three-source case. Similar results can be obtained for single-sleep and multi-sleep policies.
}

\subsubsection{LCFS-W}
\begin{figure}[htb]
\centering
\includegraphics[width=70mm]{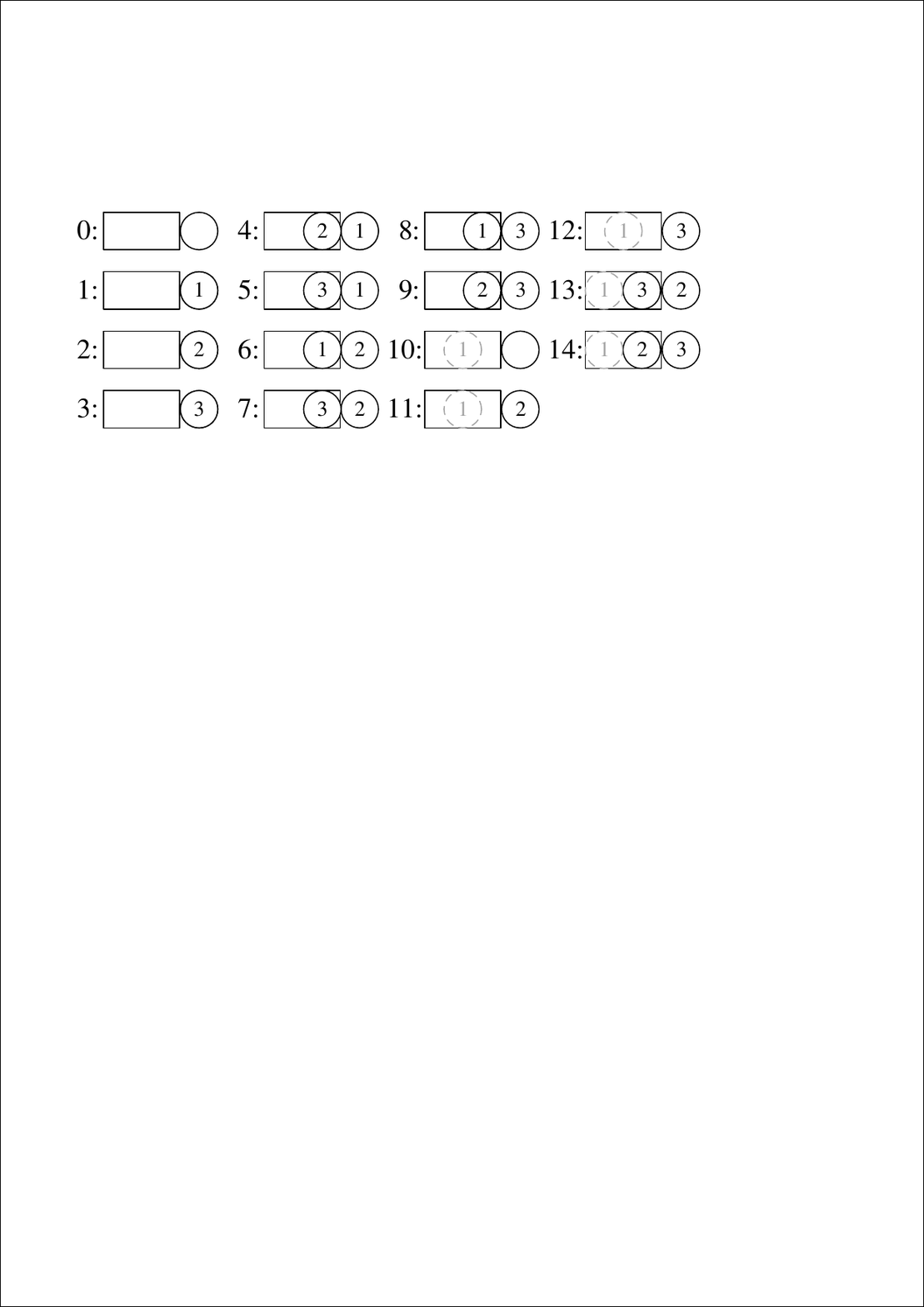}
\caption{Illustration of states with LCFS-W for 3 sources case. The rectangle on the left represents queue, the circle on the right represents the server, the solid circle represents that packets in queue, and the gray circle represents that the data from source 1 is unsynchronized.}
\label{Three source ideal LCFS-W}
\end{figure}

In LCFS-W, the discrete state is depicted in Fig. \ref{Three source ideal LCFS-W}. Different from LCFS-S, the queue holds up to one freshest packet, which is represented by a circle in the queue with a number indicating its source.

The evolution of ${\mathbf{x}}\left( t \right)$ depends on whether the system contains the packet of source 1, i.e., $b_q=0$ for $q\in\{0,2,3,7,9\}$ and $b_q = 1$ for the rest. The transitions between states can be analyzed similar to the LCFS-S case. Due to the space limitation, we ignore the detailed state transition analysis and directly present the results as follows.
\begin{equation}\label{A3}
    {{\bar \Delta_{1, \textrm{Ideal}} }} = \frac{{\sum\limits_{i = 0}^8 {{\mu ^i}{\eta _i}} }}{{\mu {\lambda _1}\left( {\mu  + \lambda_1 + \lambda_2} \right)\left( {\mu  + \lambda_1 + \lambda_3 } \right)\sum\limits_{i = 0}^7 {{\mu ^i}{\gamma _i}}}},
\end{equation} 
\begin{equation}
   \bar{P}_{\textrm{Ideal}} = \frac{{{\mu ^2}{P_\textrm{S}} + \left( {\mu \lambda  + 2{\lambda _1}{\lambda _2} + 2{\lambda _1}{\lambda _3} + 2{\lambda _2}{\lambda _3}} \right){P_\textrm{B}}}}{{\mu \lambda + 2{\lambda _1}{\lambda _2} + 2{\lambda _1}{\lambda _3} + 2{\lambda _2}{\lambda _3} + {\mu ^2}}}.
\end{equation}
where the parameters are detailed in Appendix C.

\subsubsection{LCFS-Q}\label{A4}
\begin{figure}[htb]
\centering
\includegraphics[width=70mm]{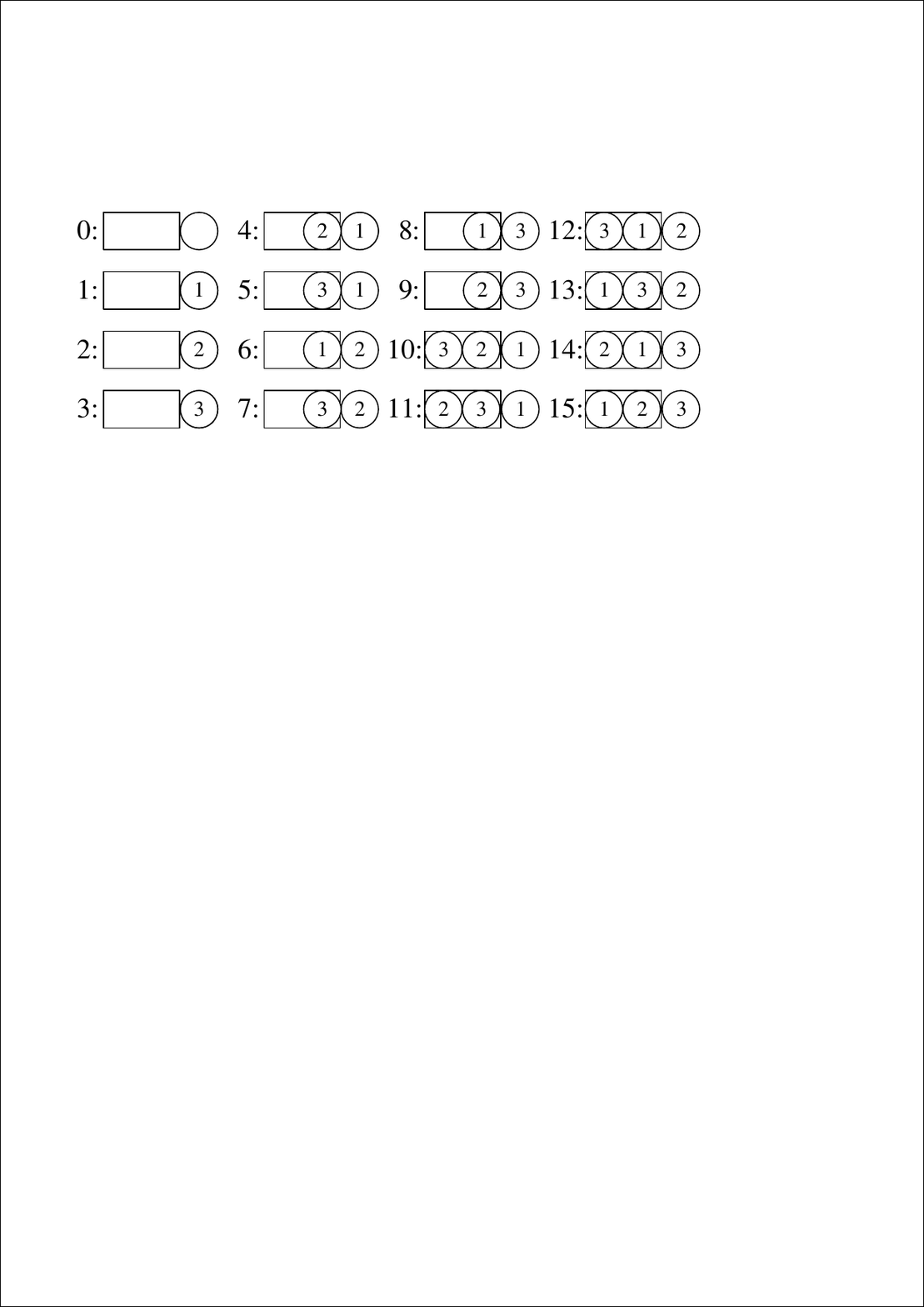}
\caption{Illustration of states with LCFS-Q for 3 sources case. The rectangle on the left represents queue, the circle on the right represents the server, the solid circle represents that packets in queue, and the gray circle represents that the data from source 1 is unsynchronized.}
\label{Three source ideal LCFS-Q}
\end{figure}

In LCFS-Q, the discrete state is depicted in Fig. \ref{Three source ideal LCFS-Q}. Different from LCFS-W, fresh packets from all the sources can be stored simultaneously. As the server can process at most one packet, the queue holds up to two packets. And the evolution of ${\mathbf{x}}\left( t \right)$ is similar to LCFS-W, i.e., $b_q=0$ for $q\in\{0,2,3,7,9\}$ and $b_q = 1$ for the rest.

The results are detailed as follows.
\begin{equation}
    {\bar \Delta_{1, \textrm{Ideal}} } = \frac{{\mu^2 {\lambda _1} + 3\mu{\lambda _1}{\lambda _2} + 3\mu{\lambda _1}{\lambda _3} + {12{\lambda _1}{\lambda _2}{\lambda _3}}}}{\mu A},
\end{equation}
\begin{equation}
    \bar{P}_{\textrm{Ideal}} = \frac{{{\mu ^3}{P_\textrm{S}} + \left( {A - {\mu ^3}} \right){P_\textrm{B}}}}{A},
\end{equation}
where $A = {\mu ^2} {\lambda}  + {\mu ^3} + 2\mu {\lambda _1}{\lambda _2} + 2\mu {\lambda _1}{\lambda _3} + 2\mu {\lambda _2}{\lambda _3} + 6{\lambda _1}{\lambda _2}{\lambda _3}$.

{For more than three sources, it is still possible to derive the closed-form expression according to the similar procedure. However, the closed-form expression is over-complex and cannot provide any insight. Thus, numerical calculation and evaluation is a necessity. The computational complexity is analyzed as follows. As each source has two discrete states, i.e., synchronized or not, the total number of system states is at the order of $O(M^2)$. Then, the computation complexity mainly lies in solving the linear equations \eqref{eq8} and \eqref{eq9}, which is of complexity $O(n^3)$ with $n$ system states. Thus, the overall computational complexity is $O(M^6)$.
}

\subsection{Arbitrary Distribution Analysis} \label{4D}
In practice, the distribution may be non-exponential. In this case, the SHS analysis can not be directly applied. However, it is still possible to analyze via SHS with some generalizations. In particular, arbitrary distribution can be represented by phase-type distribution \cite{Phase} with exponential elements. Therefore, theoretically, we can solve the problem by SHS method with any distributions. In the following, we consider two examples for a single source and 1-policy. 

\subsubsection{Constant Wake-up Time}
Assume the idle time is zero, i.e., $d=0$, and the wake-up time is a constant $\theta$. Note that the \textit{Erlang-k} distribution with mean $\frac{1}{\mu}$ and variance $\frac{1}{k\mu^2}$ can be viewed as a sum of $k$ exponentially distributed random variables with mean $\frac{1}{k\mu}$. By letting $k \to \infty$, the variance $\frac{1}{k\mu^2} \to 0$, we get a deterministic distribution, i.e., a constant.

\begin{figure}[htb]
	\centering
	\includegraphics[width=40mm]{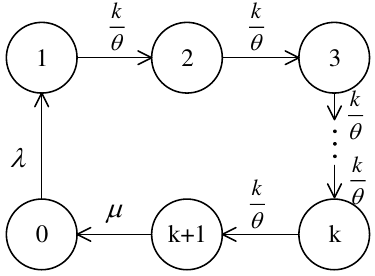}
	\caption{A constant wake-up time distribution example.}
	\label{Constant}
\end{figure}

Therefore, the Markov chain for the discrete state $q\left( t \right)$ can be estimated by the state transition in Fig. \ref{Constant} with sufficiently large $k$. The state $q=0$ indicates that there are no packets in server. State transition from $1$ to $k$ represents the server is in wake-up state and is used to simulate a constant wake-up time. The state $q=k+1$ represents the server is busy. All the transitions between states are shown in Table \ref{table:Constant}.

\begin{table}[htbp] 
\renewcommand\arraystretch{1.2}
\begin{center}
\caption{Table of Transitions for the Markov Chain in Fig. \ref{Constant}.}  
\label{table:Constant} 
\begin{tabular}{ | c | c | c | c | c | c | } 
\hline
     $l$   & ${{q_l} \to {{q'}_l}}$   & ${{\lambda ^{\left( l \right)}}}$   & ${{\mathbf{x}}{{\mathbf{A}}_l}}$   & ${{{\mathbf{A}}_l}}$   & ${{{{\mathbf{\bar v}}}_{{q_l}}}{{\mathbf{A}}_l}}$    \\ 
\hline
     $1$   & ${0 \to 1}$   & $\lambda$    & ${\left[ {{x_0}} \right]}$   & ${\left[ 1 \right]}$   & ${\left[ {{v_{0}}} \right]}$ \\
\hline
      $2$   & ${1 \to 2}$   & $\frac{k}{\theta}$    & ${\left[ {{x_0}} \right]}$   & ${\left[ 1 \right]}$   & ${\left[ {{v_{1}}} \right]}$   \\ 
\hline
      $ \vdots $   & $ \vdots $   &$ \vdots $   &$ \vdots $   &$ \vdots $   &$ \vdots $        \\ 
\hline
      $k+1$   & ${k \to k+1}$   & $\frac{k}{\theta}$    & ${\left[ {{x_0}} \right]}$   & ${\left[ 1 \right]}$   & ${\left[ {{v_{k+1}}} \right]}$   \\ 
\hline
     $k+2$   & ${k+1 \to 0}$   & $\mu$   & ${\left[ 0 \right]}$   & ${\left[ 0 \right]}$   & ${\left[ 0 \right]}$   \\
\hline
\end{tabular}
\end{center}
\end{table}

Similarly, with the SHS method, we obtain average AoS and power consumption as follows:
\begin{equation}
    {\bar \Delta } = \frac{{{{\frac{1}{\mu }}^2} + \frac{1}{\mu }\theta  + \frac{1}{2}{\theta ^2}}}{{\frac{1}{\lambda } + \frac{1}{\mu } + \theta }},
\end{equation}
\begin{equation}
    \bar{P} = \frac{{\left( {\frac{1}{\lambda } + \theta } \right){P_\textrm{S}} + \frac{1}{\mu }{P_\textrm{B}}}}{{\frac{1}{\lambda } + \frac{1}{\mu } + \theta }}.
\end{equation}

{
\subsubsection{Discrete Inter-Arrival Time}
Suppose the inter-arrival time $T$ follows a Zipf distribution 
\begin{align}
	P(T = n) = \frac{\sigma}{n^s}, \quad n =1, 2, \cdots, N,
\end{align}
where $s \ge 0 $ is the skew parameter, and $\sigma$ is the normalization factor. For simplicity, we set $N = 2$. Thus, we have 
\begin{align}
	P(T = 1) &= \sigma,\\P(T = 2) &= 1-\sigma,
\end{align}
where $\sigma = \left(1+\frac{1}{2^s}\right)^{-1}.$ It can be represented by a $2k$-phases Markov chain as $k \rightarrow \infty$ with transition intensity $k$ from state $i$ to $i+1$ for $i \neq k$, with intensity $(1-\sigma)k$ from $k$ to $k+1$, and with intensity $\sigma k$ from $k$ to $0$. Consider the ideal sleep model with $\theta=0$ and $d=0$, the state transition is shown in Fig.~\ref{fig:phase_trans}, where $0, \cdots, 2k$ refers to arrival phases in sleep mode, $0'$ denotes an arrival and $1', \cdots, 2k'$ refers to arrival phases in busy mode.

\begin{figure}[htbp]
	\centering
	\includegraphics[width=3.2in]{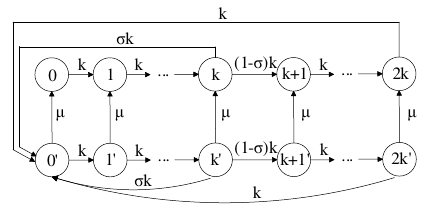}
	\caption{A discrete inter-arrival time distribution example.}
	\label{fig:phase_trans}
\end{figure}

With the SHS method, and then by letting $k\rightarrow \infty$, we can obtain the average AoS and power consumption for Zipf distributed inter-arrival time as follows
\begin{align}
	\bar{\Delta} &= \frac{\sigma e^{-\mu} + 2(1-\sigma) e^{-2\mu}}{1-\sigma e^{-\mu} - (1-\sigma) e^{-2\mu}} \cdot \frac{(1-e^{-\mu})(1+(1-\sigma)e^{-\mu})}{(2-\sigma)\mu}	\nonumber\\
	& + \frac{1 + (\mu-1) e^{-\mu}}{(2-\sigma)\mu^2}  + \frac{(1-\sigma)((\mu+1) e^{-\mu} - (2\mu + 1)e^{-2\mu})}{(2-\sigma) \mu^2},	
\end{align}
\begin{align}
	\bar{P} &= \frac{\mu - 1 + e^{-\mu} + (1-\sigma)(\mu - e^{-\mu} + e^{-2\mu})}{(2-\sigma)\mu} P_{\mathrm S} \nonumber\\
	&\quad + \frac{1-e^{-\mu} + (1-\sigma)(e^{-\mu} - e^{-2\mu})}{(2-\sigma)\mu} P_{\mathrm B}. 
\end{align}

}

% \begin{figure}[t]
% %\centeringp
% \includegraphics[height=5cm ,width=78mm]{Figure11-eps-converted-to.eps}
% \caption{Network scenario with clustering.}
% \label{Figure11}
% \end{figure}
% \begin{figure}[t]
% %\centering
% \includegraphics[width=88mm]{Figure22-eps-converted-to.eps}
% \caption{Intra-cluster access period and inter-cluster access period.}
% \label{Figure22}

% \end{figure}

% \begin{figure*}[t]
% \centering
% \subfigure[]{\includegraphics[width=3.45in,height=2.3in]{Figure0-eps-converted-to.eps}}
% \subfigure[]{\includegraphics[width=3.45in,height=2.3in]{Figure0b-eps-converted-to.eps}}
% %\includegraphics[widtph=78mm]{Fpigure0.eps}p
% \caption{Probability of successful access $p_{cluster}$ versus number of clusters $C$ and length of intra-cluster access period $t$, $\mu = \tfrac{1}{3}$, $T = 20$, $M = 20$, $N = 10000$. (a) $p_{cluster}$ versus $C$, $t\in \left\{ 4,8,12,16 \right\}.$ (b) $p_{cluster}$ versus $t$, $C\in \left\{ 100,150,200,250 \right\}.$}
% \label{Figure0}
% \vspace{-0.1cm}
% \end{figure*}%p

% \begin{equation}
%     p_{intra}^{} = {\left( {1 - \tfrac{1}{{M \cdot \mu  \cdot t}}} \right)^{\tfrac{N}{C} - 1}}.
%     \label{eq2}
% \end{equation}%

% \begin{equation}
% \begin{aligned}
%     p_{cluster} &= p_{intra}^{} \cdot p_{inter}^{} \\
%     &= {\left( {1 - \tfrac{1}{{M \cdot \mu  \cdot t}}} \right)^{\tfrac{N}{C} - 1}} \cdot {\left( {1 - \tfrac{1}{{M \cdot (T - t)}}} \right)^{C - 1}}.
%     \label{eq5}
% \end{aligned}
% \end{equation}%p

\begin{figure*}[t]
	\centering
	\subfigure[$\mu = 1$, $\lambda = 0.5$]{\includegraphics[width=2.35in]{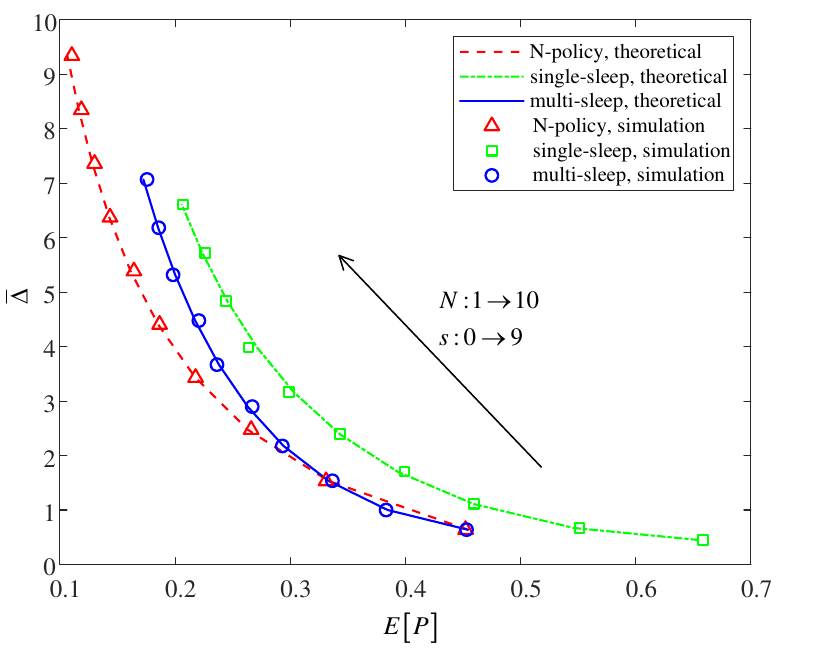}}
	\subfigure[$\mu = 1$, $\lambda = 1$]{\includegraphics[width=2.35in]{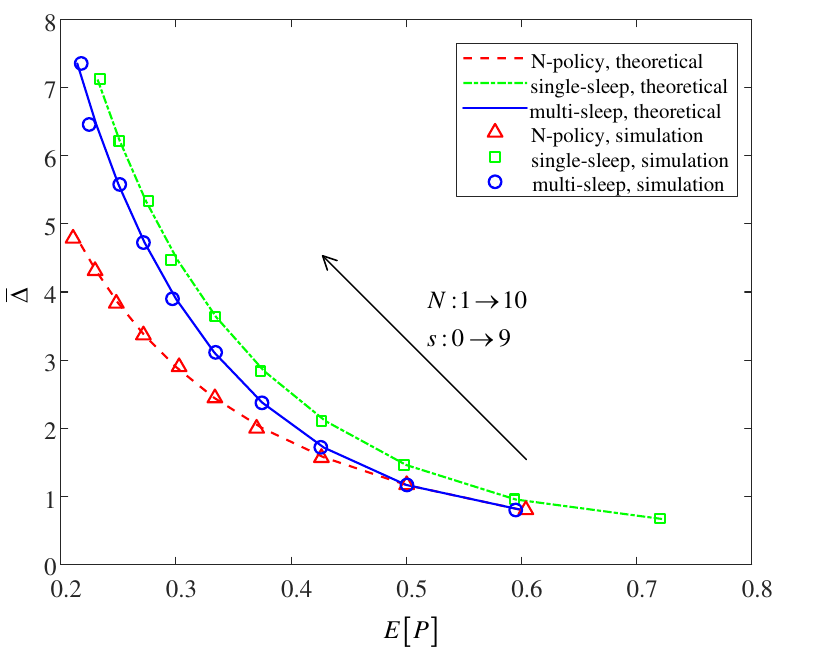}}
	\subfigure[$\mu = 1$, $\lambda = 2$]{\includegraphics[width=2.35in]{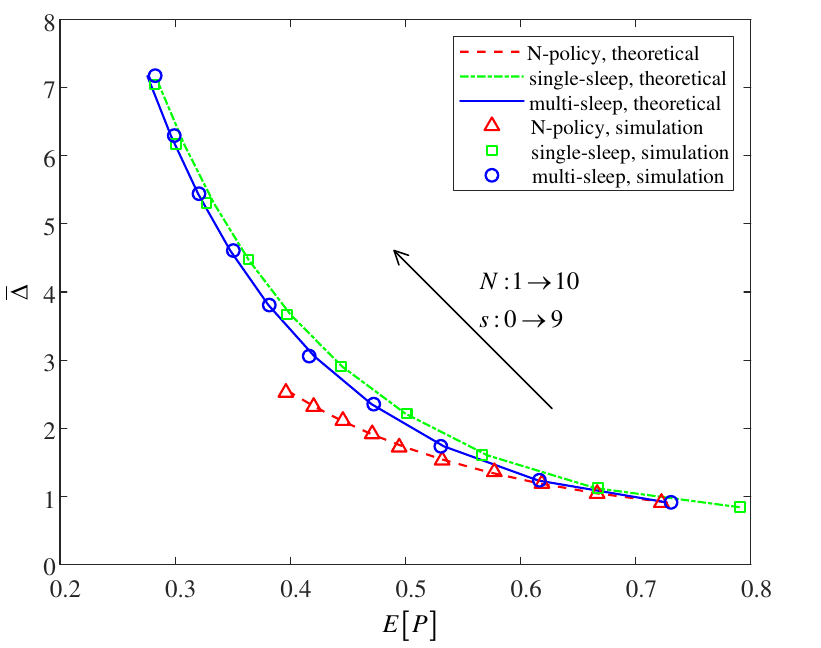}}
	\caption{The impact of $N$ and $s$ on the age-energy trade-off in single source case with $d = \theta =1$.}
	\label{fig_one_source}
\end{figure*}

\section{Numerical Results}\label{V}
In this section, we show the trade-off between the average AoS and the average power consumption through numerical simulations. We set the power consumed in each state in the sleep model as ${P_{\textrm{B}}}=1$, ${P_{\textrm{I}}}={P_{\textrm{W}}}=0.5$, ${P_{\textrm{S}}}=0$.

\subsection{Single Source Simulation} 
First, we consider the single source case. Fig. \ref{fig_one_source} illustrates the impact of the wake-up policies on the trade-off performance. Firstly, it can be seen that the simulation curves well meet the analytical ones, which validates our theoretical analysis. We can observe that with the increase of sleep parameters ($N$ and $s$), the average power consumption decreases while the average AoS increases across all policies. This is due to the direct effect of increasing the sleeping time. Comparison of the three wake-up policies reveals that the N-policy achieves the minimum AoS for a given power consumption. We also observe that the trade-off range of the N-policy shrinks as the arrival rate $\lambda$ increases. This is because when the arrival rate is low, the server sleeps for a long time under the N-policy with large $N$, resulting in very large AoS and very small power consumption. Finally, we observe that with the increase of the arrival rate $\lambda$, the gap between the single-sleep and the multi-sleep becomes small. This is because with the fast arrival of packets, it is unlikely that no packet will arrive within the first sleep period. Thus, in the multi-sleep policy, the server is woken up after one sleep period most of the time, which is equivalent to single-sleep.

\begin{figure}[!t]
\centering
\subfigure[Impact of $d$ with $\theta=1$.]{\includegraphics[width=2.5in]{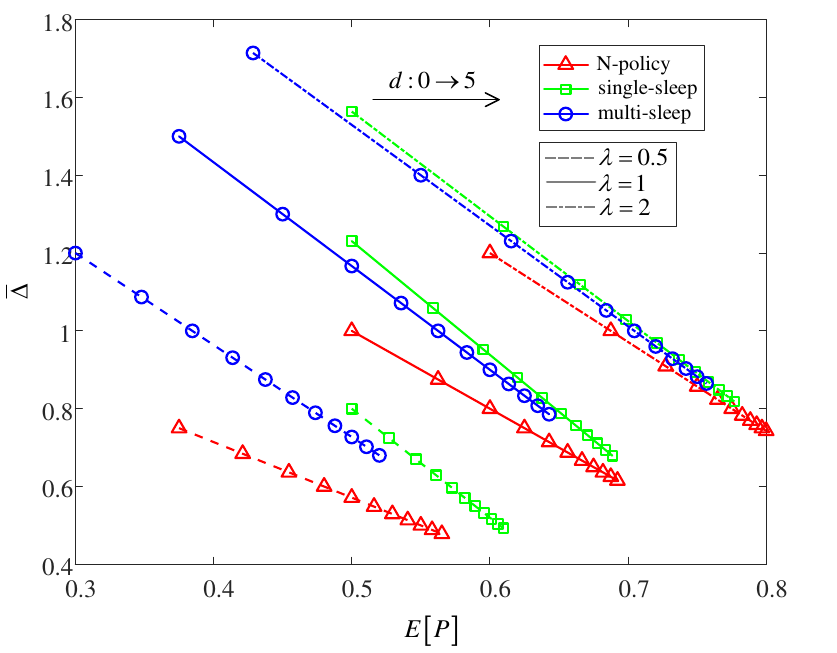}\label{fig_one_source_d}}\\
\subfigure[Impact of $\theta$ with $d=1$.]{\includegraphics[width=2.5in]{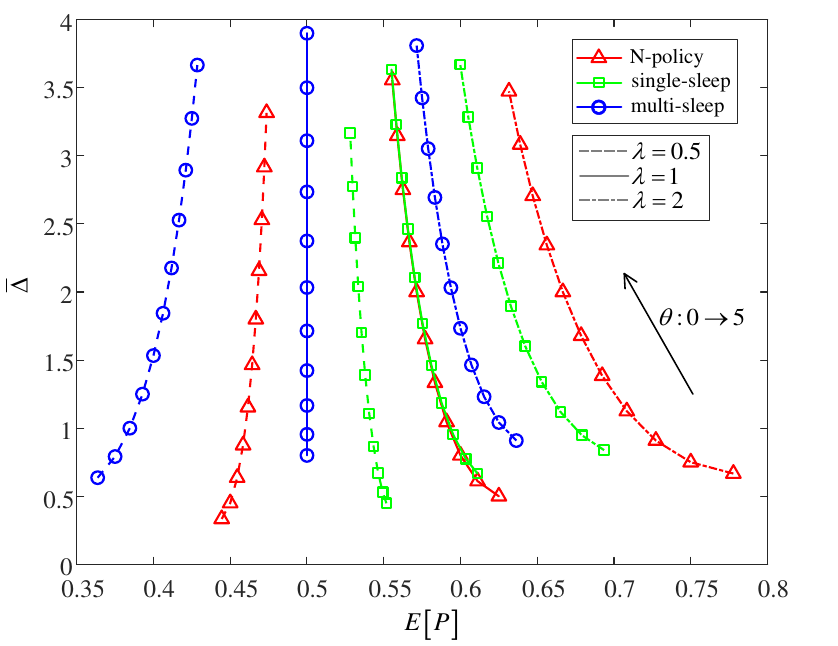}\label{fig_one_source_theta}}
\caption{The impact of $d$ and $\theta$ on the age-energy trade-off in single source case with $\mu = 1$, $N = s =1$.}
\label{fig_one_source_d_theta}
\end{figure}

% \begin{figure}[htb]
% \centering
% \includegraphics[width=85mm]{fig_one_source_d.pdf}
% \caption{The impact of $d$ on the age-energy trade-off in single source case with $\mu = 1$, $N = s =1$.}
% \label{fig_one_source_d}
% \end{figure}
% \begin{figure}[htb]
% \centering
% \includegraphics[width=85mm]{fig_one_source_theta.pdf}
% \caption{The impact of $\theta$ on the age-energy trade-off in single source case with $\mu = 1$, $N = s =1$.}
% \label{fig_one_source_theta}
% \end{figure}

Fig.~\ref{fig_one_source_d_theta} illustrates the trade-off performance of different wake-up policies by varying system parameters $d$ and $\theta$. It can be observed from Fig.~\ref{fig_one_source_d} that with the increase of idle time $d$, the average power consumption increases and the average AoS decreases. This is because with longer idle time, the server is more likely to enter busy state directly to process packets, rather than sleep. It is also seen from Fig.~\ref{fig_one_source_theta} that the average AoS increases as the wake-up time $\theta$ increases, because the wake-up time slows down the processing of packets. However, the average power consumption exhibits different trends for different cases. When the system load is low, i.e., $\lambda < \mu$, the server turns to sleep state very frequently. As $\theta$ increases, the time that server spends in the wake-up state also increases, leading to higher power consumption. In contrast, when $\lambda > \mu$, the server remains busy most of the time. Hence, increasing the low-power wake-up time in this case reduces the average power consumption.

\subsection{Two Sources Simulation} 
\begin{figure*}[htb]
\centering
\subfigure[AoS of source 1 versus power. \newline ~~~$\lambda_2=1$, $s = 0.25$.]{\includegraphics[width=2.3in]{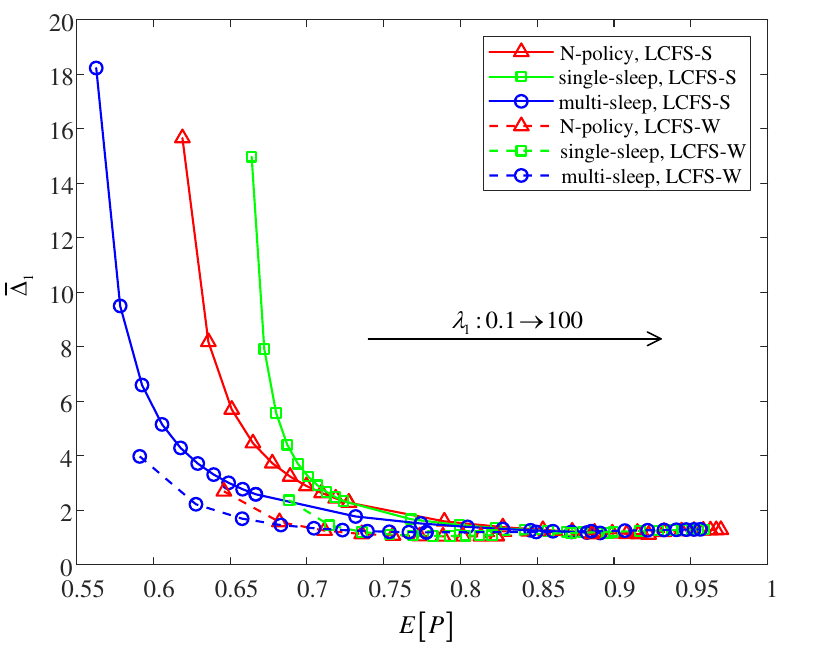}\label{fig_two_source_lambda1}}
\subfigure[AoS of source 2 versus power. \newline ~~~$\lambda_2=1$, $s = 0.25$.]{\includegraphics[width=2.3in]{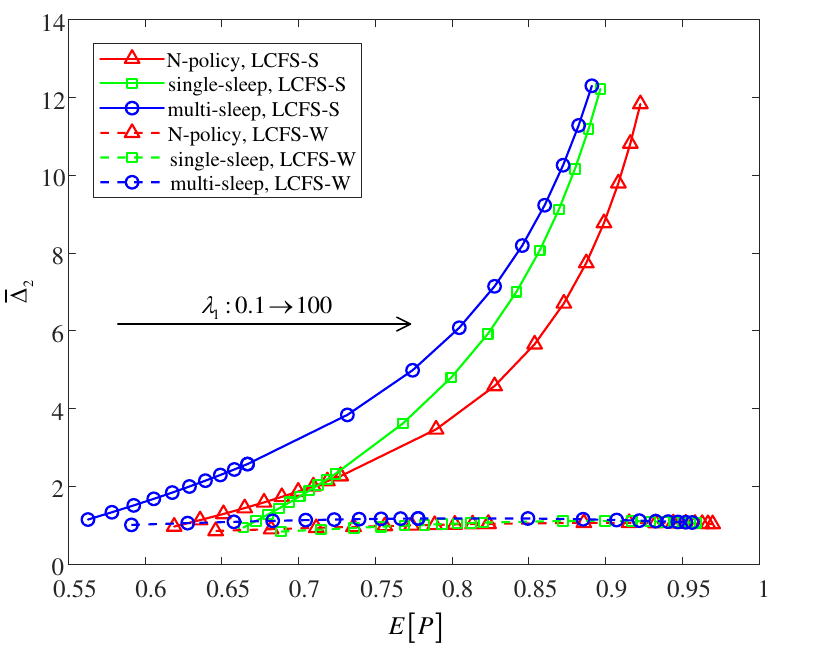}\label{fig_two_source_lambda2}}
\subfigure[Average sum AoS versus power. \newline ~~~$\lambda_1+\lambda_2=2$, $s = 1$.]{\includegraphics[width=2.4in]{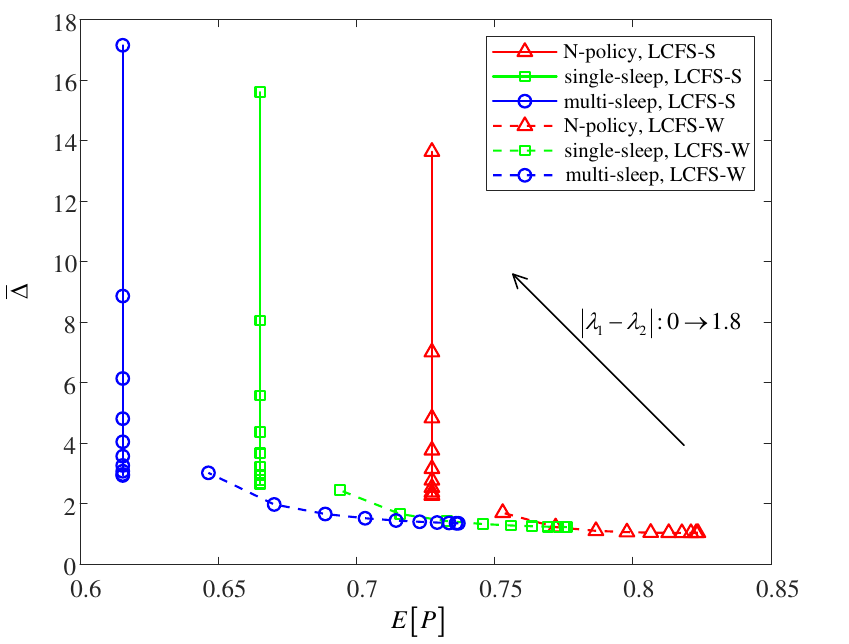}\label{fig_two_source_diff}}
\caption{Age-Energy trade-off in two sources case. $\mu = 1$, $d = 1$, $\theta =1$, $N = 1$.}
\label{fig_two_source}
\end{figure*}
Secondly, we analyze the case of the two sources. Figs.~\ref{fig_two_source_lambda1} and \ref{fig_two_source_lambda2} show the age-energy trade-off of source 1 and source 2 with fixed server processing rate $\mu=1$ and packet arrival rate of source 2 $\lambda_2=1 $. It can be seen that when $\lambda_1 $ increases, the average power consumption of all curves increases, while the average AoS of source 1 decreases. This observation aligns with intuition since a higher arrival rate means the server spends more time processing packets, leading to higher power consumption. Additionally, as the server is occupied more often by the packets from source 1 with a larger $\lambda_1$, the AoS of source 1 decreases. 
The two preemption strategies show significant difference in AoS. Under the LCFS-S policy, a packet from source 1 may be preempted by a packet from source 2 in the service. Hence, when $\lambda_1$ is much less than $\lambda_2$, the packet from source 1 is difficult to be served, resulting in a larger age, and vice versa. In contrast, preemption in the service between sources does not occur under the LCFS-W policy, and once the server is occupied by a source, it will certainly synchronize. Therefore, both sources achieve lower AoS in LCFS-W compared with LCFS-S.
% \begin{figure}[htb]
% \centering
% \includegraphics[width=85mm]{fig_two_source.pdf}
% \caption{Age-Energy trade-off versus packets arrival rate of source 1 $\lambda_1 = \left[0.1, 100\right]$ in two sources case, with $\mu = 1$, $\lambda_2=1$ $d = \theta =1$, $s = 0.25$, $N = 1$.}
% \label{fig_two_source_lambda}
% \end{figure}

% \begin{figure}[htb]
% \centering
% \includegraphics[width=85mm]{fig_two_source_diff.pdf}
% \caption{Age-Energy trade-off versus difference of packets arrival rates between two sources $\left| {{\lambda _1} - {\lambda _2}} \right| = \left[0, 1.8\right]$, with $\mu = 1$, $\lambda_1+\lambda_2=2$, $d = \theta =1$, $s = 1$, $N = 1$.}
% \label{fig_two_source_diff}
% \end{figure}

Furthermore, fig.~\ref{fig_two_source_diff} illustrates the system sum AoS $\bar{\Delta} = {{\bar{\Delta}} _ {1}} + {{\bar{\Delta}} _ {2}}$ versus the average power consumption by varying difference of the two sources' arrival rates $|\lambda_1 - \lambda_2|$ with fixed total arrival rate $\lambda=\lambda_1+\lambda_2 = 2$. As seen in this figure, the sum AoS under LCFS-S policy increases sharply as the difference increases, while the increase of LCFS-W is very slow, which is consistent with the conclusion of Figs.~\ref{fig_two_source_lambda1} and \ref{fig_two_source_lambda2}. On the other hand, the average power consumption of LCFS-S does not change with the difference. This is because the power consumption with LCFS-S is only related to the total arrival rate $\lambda$.
When the difference of the arrival rates is significant, the LCFS-W policy is preferred as it can greatly reduce sum AoS while slightly increasing power consumption. When the difference is marginal, on the contrary, the LCFS-S policy is more promising as it can greatly reduce power consumption while slightly increasing sum AoS.

\subsection{Three and More Sources} 
\begin{figure*}[htb]
\centering
\subfigure[AoS of source 1 versus power. \newline $\lambda_2=\lambda_3=1$.] {\includegraphics[width=2.32in]{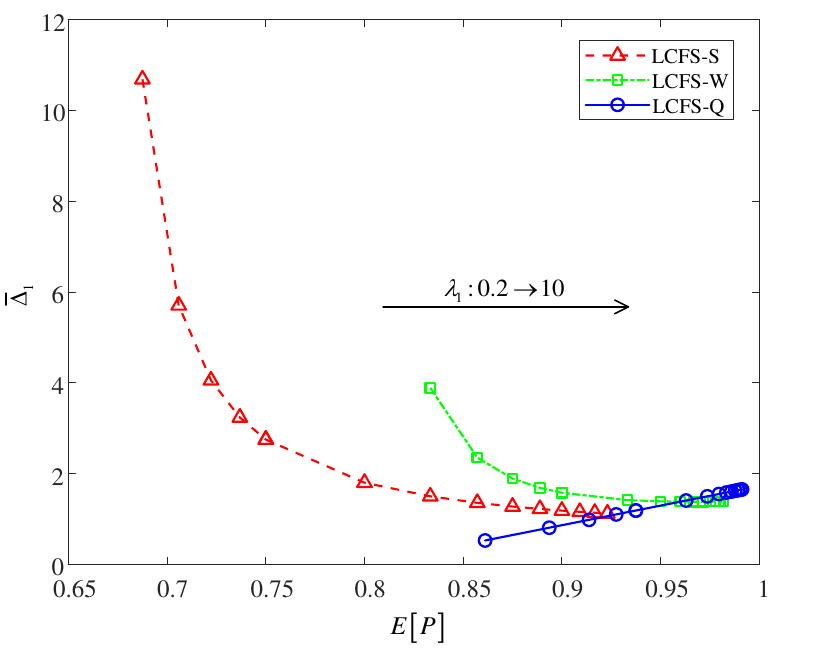}\label{fig_three_source_lambda1}}
\subfigure[AoS of source 2 (3) versus power. \newline $\lambda_2=\lambda_3=1$.] {\includegraphics[width=2.32in]{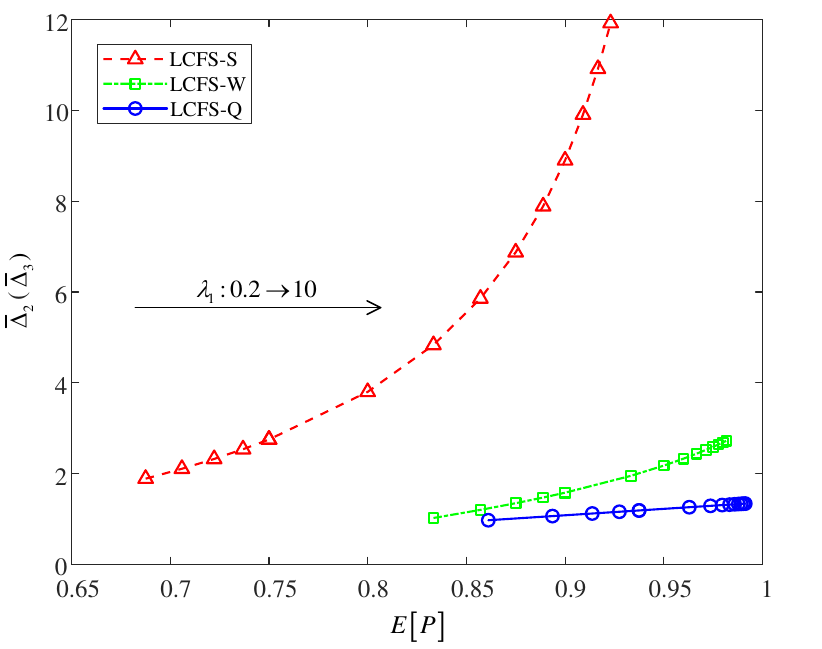}\label{fig_three_source_lambda2}}
\subfigure[Average sum AoS versus power. \newline $\lambda_1+\lambda_2=2$, $\lambda_3=1$.] {\includegraphics[width=2.32in]{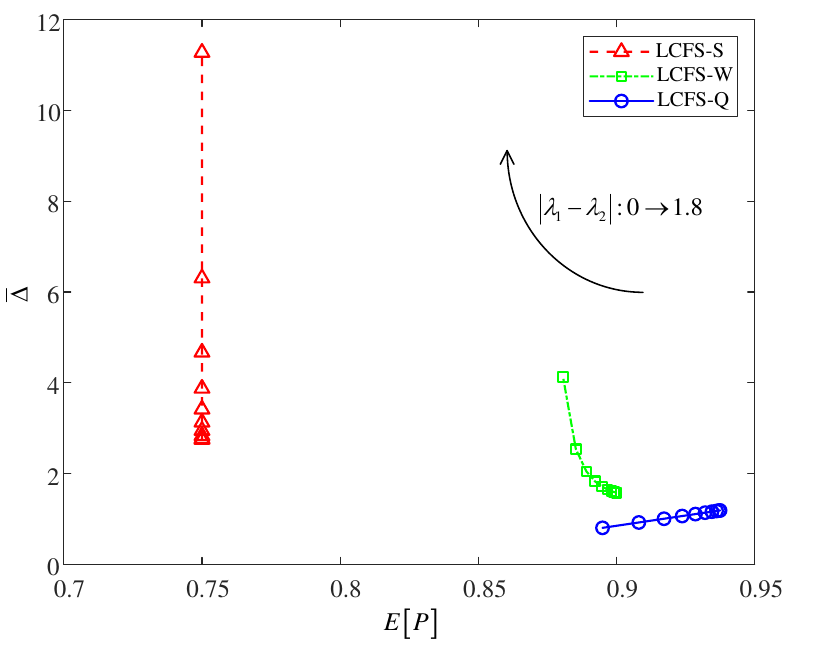}\label{fig_three_source_diff}}
\caption{Age-Energy trade-off in three sources case with ideal sleep model. $\mu = 1$.}
\label{fig_three_source}
\end{figure*}

Next, we consider three sources with 1-policy. Similarly, we show the age-energy performance of source 1 and source 2 (source 3) by varying $\lambda_1 $ with fixed server processing rate $\mu=1$ and packet arrival rate of source 2 and source 3 $\lambda_2=\lambda_3=1$ in Figs.~\ref{fig_three_source_lambda1} and \ref{fig_three_source_lambda2}. Similar to the two-sources case, the average AoS of source 1 with LCFS-S and LCFS-W decreases as $\lambda_1$ increases. Interestingly, the average AoS of source 1 with LCFS-Q increases with the increase of $\lambda_1$. The reason is that high arrival rate results in non-empty buffer with high probability, which causes the increase of AoS due to queuing delay. We also depict the sum average AoS versus the average power in Fig.~\ref{fig_three_source_diff}. For convenience, we fix $\lambda_3$ and vary the difference between $\lambda_1$ and $\lambda_2$. As seen in this figure, the trends of LCFS-S and LCFS-W are similar to Fig. \ref{fig_two_source_diff}. For LCFS-Q, both the total average AoS and average power consumption decrease with the increase of arrival rate difference. It indicates that when the total arrival rate is constant, the greater the difference between sources, the better the effect of LCFS-Q.

When compared to the three preemption strategies, LCFS-S exhibits low average power consumption, but has very high total average AoS. LCFS-Q, on the other hand, has a very small total average AoS, but high average power consumption. LCFS-W performs neutral, as it retains the packet preemption mechanism but limits preemption to the queue, which ensures that data packets being serviced are completely processed. Therefore, when the system requires high energy efficiency, LCFS-S strategy should be preferred. When the system requires greater information synchronization, it is more appropriate to adopt LCFS-Q strategy.

\begin{figure}[!t]
	\centering
	\subfigure[Average AoS of source 1.]{\includegraphics[width=2.5in]{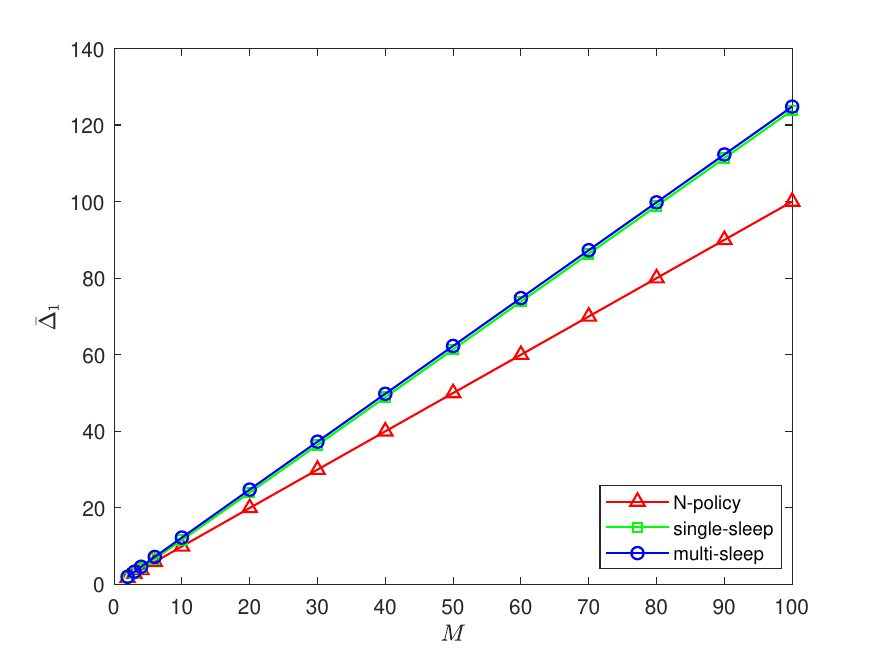}\label{fig_agevsM}}\\
	\subfigure[Average power consumption.]{\includegraphics[width=2.5in]{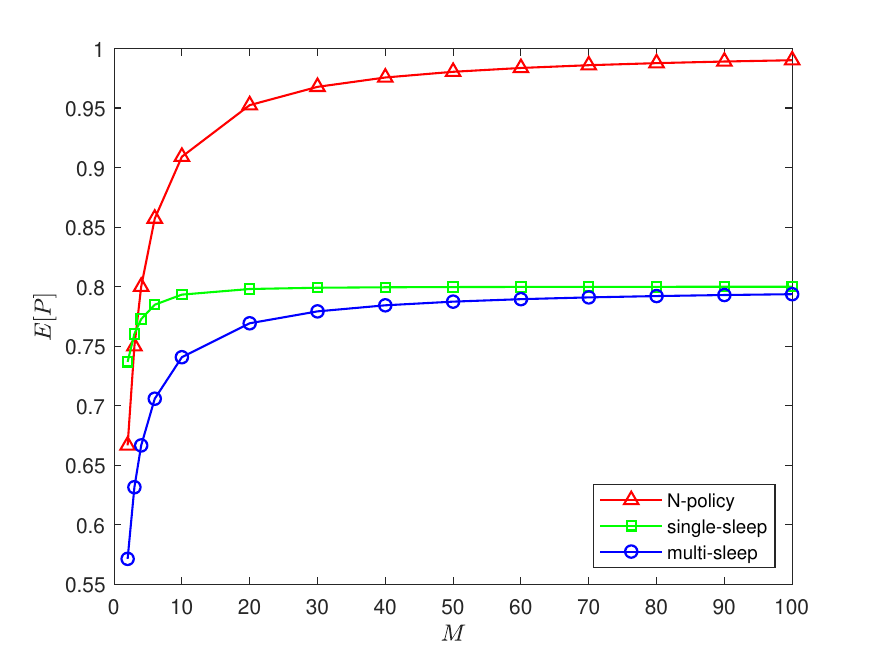}\label{fig_powervsM}}
	\caption{The impact of the number of sources $M$ on the age-energy performance with LCFS-S policy. $\lambda_1 = \cdots = \lambda_M = 1, \mu = 1, N = 1, s = 0.25$.}
	\label{fig_multisource}
\end{figure}

{Finally, we evaluate the impact of the number of sources $M$ on the performance with LCFS-S policy for ideal sleep model, as shown in Fig.~\ref{fig_multisource}. It can be seen that as the number of sources increases, both average AoS and power consumption increase. When the number of sources is large, the gap of average AoS between 1-policy and the other two policies increases at the cost of increasing gap of power consumption. The reason is that 1-policy tends to wake up much quicker than the others when the total arrival rate is large. It is also seen that the power consumption of multi-sleep gradually converges to that of single sleep, which keeps constant. This is because there is almost always packets waiting to be processed after the first sleep due to high arrival rate.
}
\section{Conclusion}\label{VI}
In this paper, we derive explicit expressions of the average AoS and average power consumption for a multi-source single-server system with different wake-up policies and preemption strategies. Numerical results show that changing the parameters of the wake-up policies can enable a trade-off between power consumption and AoS. With the same settings, the N-policy has the best trade-off performance. The trade-off range of the N-policy widens as the arrival rate decreases. When the arrival rate is high, single-sleep and multi-sleep tend to have similar trade-off performance. Additionally, different preemption strategies have distinct advantages in different situations. The LCFS-S strategy should be adopted when the system requires high energy savings or the arrival rate difference between sources is small. When the system requires high information freshness or the arrival rate difference between sources is large, the LCFS-Q strategy is more appropriate. LCFS-W combines the strengths and weaknesses of the two strategies. In future work, it would be interesting to extend the analysis to broadcast networks or multi-hop networks. {When giving a specific information structure of the sources, Age of Incorrect Information (AoII) \cite{aoii} can be further considered as a more generalized performance metric which measures not only the time difference since out-of-sync, but also the estimation error of the data.}

\newpage
{~}
{

	\newpage
	\appendices
	\section{Parameters in (\ref{A1})}\label{AA}
	\begin{align*}
		{\gamma _0} &{=} s{\mu ^2}\left( {2{\lambda _2} + 3{\lambda _1}} \right) + \mu \left( {2{\lambda _2} + 2{\lambda _1} + \mu } \right), \\ 
		{\gamma _1} &{=} {s^2}{\mu ^2}\left( {3{\lambda _2} + 4{\lambda _1}} \right) + 2s\mu \left( {2{\lambda _2} + 3{\lambda _1}} \right), \\ 
		{\gamma _2} &{=} {s^3}{\mu ^2}\left( {2{\lambda _2} + 3{\lambda _1}} \right) + {s^2}\mu \left( {4{\lambda _2} + 7{\lambda _1}} \right) + s\left( {2{\lambda _2} + 3{\lambda _1}} \right) {+} 1, \\ 
		{\gamma _3} &{=} 2{s^3}\mu \left( {{\lambda _2} + 2{\lambda _1}} \right) + {s^2}\left( {{\lambda _2} + 3{\lambda _1}} \right), \\ 
		{\gamma _4} &{=} {s^4}{\mu ^2} + {s^4}\mu {\lambda _1} + {s^3}{\lambda _1},\\
		{\eta _0} &{=} s{\mu ^3}{\lambda _2}\left( {3{\lambda _2} {+} 4{\lambda _1}} \right) {+} {\mu ^3}{\lambda _2}, \\ 
		{\eta _1} &{=} {s^3}{\mu ^3}\left( {11\lambda _2^3 {+} 30{\lambda _1}\lambda _2^2 {+} 23\lambda _1^2{\lambda _2} {+} 5\lambda _1^3} \right) \\ 
		&  \quad {+} {s^2}{\mu ^3}\left( {8\lambda _2^2 {+} 10{\lambda _1}{\lambda _2} {+} \lambda _1^2} \right) {+} {s^3}{\mu ^4}\left( {2\lambda _2^2 {+} 4{\lambda _1}{\lambda _2} {+} \lambda _1^2} \right) \\ 
		&  \quad {+} s{\mu ^2}\left( {3{\lambda _2} {+} {\lambda _1}} \right)\left( {3{\lambda _2} {+} 4{\lambda _1}} \right) {+} {\mu ^2}\left( {3{\lambda _2} {+} {\lambda _1}} \right) {+} {s^2}{\mu ^4}{\lambda _2},  \\
		{\eta _2} &{=} {s^4}{\mu ^3}\left( {10\lambda _2^3 {+} 30{\lambda _1}\lambda _2^2 {+} 27\lambda _1^2{\lambda _2} {+} 9\lambda _1^3} \right) \\ 
		&  \quad {+} {s^2}{\mu ^2}\left( {16\lambda _2^2 {+} 29{\lambda _1}{\lambda _2} {+} 8\lambda _1^2} \right) {+} \mu \left( {3{\lambda _2} {+} 2{\lambda _1}} \right) \\ 
		&  \quad {+} s\mu \left( {3{\lambda _2} {+} 2{\lambda _1}} \right)\left( {3{\lambda _2} {+} 4{\lambda _1}} \right) {+} {s^4}{\mu ^4}\left( {3\lambda _2^2 {+} 4{\lambda _1}{\lambda _2} {+} 3\lambda _1^2} \right),  \\
		{\eta _3} &{=} {s^3}{\mu ^2}\left( {17\lambda _2^2 {+} 36{\lambda _1}{\lambda _2} {+} 11\lambda _1^2} \right) {+} {s^2}\mu \left( {12\lambda _2^2 {+} 28{\lambda _1}{\lambda _2} {+} 13\lambda _1^2} \right), \\ 
		{\eta _4} &{=} {s^5}{\mu ^4}\left( {2{\lambda _2} {+} 3{\lambda _1}} \right) {+} 2{s^4}{\mu ^2}\left( {5\lambda _2^2 {+} 13{\lambda _1}{\lambda _2} {+} 5\lambda _1^2} \right) \\ 
		&  \quad {+} {s^3}\mu \left( {9\lambda _2^2 {+} 24{\lambda _1}{\lambda _2} {+} 11\lambda _1^2} \right) {+} s\left( {3{\lambda _2} {+} 4{\lambda _1}} \right) {+} 1, \\ 
		{\eta _5} &{=} {s^5}{\mu ^3}\left( {5{\lambda _2} {+} 7{\lambda _1}} \right) {+} {s^5}{\mu ^2}\left( {3\lambda _2^2 {+} 9{\lambda _1}{\lambda _2} {+} 5\lambda _1^2} \right) \\ 
		&  \quad {+} {s^4}\mu \left( {3\lambda _2^2 {+} 11{\lambda _1}{\lambda _2} {+} 5\lambda _1^2} \right) {+} 3{s^2}\left( {{\lambda _2} {+} 2{\lambda _1}} \right), \\ 
		{\eta _6} &{=} {s^6}{\mu ^4} {+} {s^6}{\mu ^3}\left( {{\lambda _2} {+} 2{\lambda _1}} \right) {+} {s^5}\mu {\lambda _1}\left( {2{\lambda _2} {+} {\lambda _1}} \right) {+} {s^3}\left( {{\lambda _2} {+} 4{\lambda _1}} \right), \\ 
		{\eta _7} &{=} {s^6}{\mu ^2}{\lambda _1} {+} {s^4}{\lambda _1}, \\ 
	\end{align*}
\newpage
	
	\section{Parameters in (\ref{A2})}\label{AB}
	\begin{align*}
		A &= s\mu \lambda  {+} \mu  {+} \lambda,  \\ 
		{\eta _0} &= s{\mu ^4}{\lambda _2} {+} {\mu ^3}{\lambda _2}, \\ 
		{\eta _1} &= {s^2}{\mu ^2}\left( {6\lambda _1^3 {+} 27\lambda _1^2{\lambda _2} {+} 30{\lambda _1}\lambda _2^2 {+} 10\lambda _2^3} \right)  \\
		& \quad {+}2\left( {{s^2}{\mu ^3} {+} s{\mu ^2}} \right)\left( {2{\lambda _1}^2 {+} 8{\lambda _1}{\lambda _2} {+} 5{\lambda _2}^2} \right) \\ 
		& \quad {+}\left( {{s^2}{\mu ^4} {+} {\mu ^2}} \right)\left( {{\lambda _1} {+} 3{\lambda _2}} \right) {+} s{\mu ^3}\left( {{\lambda _1} {+} {\lambda _2}} \right),\\
		{\eta _2} &= \left( {{s^3}{\mu ^3} {+} s\mu } \right)\left( {5{\lambda _1}^2 {+} 12{\lambda _1}{\lambda _2} {+} 6{\lambda _2}^2} \right) \\
		& \quad {+}\left( {{s^3}{\mu ^4} {+} \mu } \right)\left( {2{\lambda _1} {+} 3{\lambda _2}} \right), \\ 
		{\eta _3} &= \left( {{s^3}{\mu ^2} {+} {s^2}\mu } \right)\left( {4{\lambda _1}^2 {+} 9{\lambda _1}{\lambda _2} {+} 3{\lambda _2}^2} \right), \\
		{\eta _4} &= {s^4}{\mu ^4} {+} \left( {{s^4}{\mu ^3} {+} s} \right)\left( {2{\lambda _1} {+} {\lambda _2}} \right) {+} {s^3}\mu \left( {{\lambda _1} {+} 2{\lambda _2}} \right) {+} 1, \\ 
		{\eta _5} &= {s^4}{\mu ^2}{\lambda _1} {+} {s^2}{\lambda _1}. \\ 
	\end{align*} 
	
	\begin{figure*} %hb代表放在文章底部，%ht为放在文章顶部 
		\centering
		\section*{Appendix C \\Parameters in (\ref{A3})}\label{AC}
		\begin{align*}
			{\eta _0} &= \lambda_{1}\,(\lambda_{1}+\lambda_{2})^2\,(\lambda_{3}+\lambda_{1})^2\,\lambda\,(3\,\lambda_{1}^2\,\lambda_{3}^2+7\,\lambda_{2}\,\lambda_{1}\,\lambda_{3}^2+4\,\lambda_{2}^2\,\lambda_{3}^2+3\,\lambda_{1}^3\,\lambda_{3}+12\,\lambda_{2}\,\lambda_{1}^2\,\lambda_{3}+7 \,\lambda_{2}^2\,\lambda_{1}\,\lambda_{3}+3\,\lambda_{2}\,\lambda_{1}^3+3\,\lambda_{2}^2\,\lambda_{1}^2),\\
			{\eta _1} &= \left(\lambda_{1}+\lambda_{2}\right)\,\left(\lambda_{3}+\lambda_{1}\right)\,(9\,\lambda_{1}^3\,\lambda_{3}^4+29\,\lambda_{2}\,\lambda_{1}^2\,\lambda_{3}^4+28\,\lambda_{2}^2\,\lambda_{1}\,\lambda_{3}^4+8\,\lambda_{2}^3\,\lambda_{3}^4+40\,\lambda_{1}^4\,\lambda_{3}^3+149\,\lambda_{2}\,\lambda_{1}^3\,\lambda_{3}^3+171\,\lambda_{2}^2\,\lambda_{1}^2\,\lambda_{3}^3 \\
			&  \quad+72\,\lambda_{2}^3\,\lambda_{1}\,\lambda_{3}^3+8\,\lambda_{2}^4\,\lambda_{3}^3+54\,\lambda_{1}^5\,\lambda_{3}^2+247\,\lambda_{2}\,\lambda_{1}^4\,\lambda_{3}^2+334\,\lambda_{2}^2\,\lambda_{1}^3\,\lambda_{3}^2+171\,\lambda_{2}^3\,\lambda_{1}^2\,\lambda_{3}^2+28\,\lambda_{2}^4\,\lambda_{1}\,\lambda_{3}^2+24\,\lambda_{1}^6\,\lambda_{3} \\
			&  \quad+151\,\lambda_{2}\,\lambda_{1}^5\,\lambda_{3}+247\,\lambda_{2}^2\,\lambda_{1}^4\,\lambda_{3}+149\,\lambda_{2}^3\,\lambda_{1}^3\,\lambda_{3}+29\,\lambda_{2}^4\,\lambda_{1}^2\,\lambda_{3}+\lambda_{1}^7+24\,\lambda_{2}\,\lambda_{1}^6+54\,\lambda_{2}^2\,\lambda_{1}^5+40\,\lambda_{2}^3\,\lambda_{1}^4+9\,\lambda_{2}^4\,\lambda_{1}^3), \\
			{\eta _2} &= 16\,\lambda_{2}^3\,\lambda_{3}^5+42\,\lambda_{1}\,\lambda_{2}^2\,\lambda_{3}^5+34\,\lambda_{1}^2\,\lambda_{2}\,\lambda_{3}^5+9\,\lambda_{1}^3\,\lambda_{3}^5+40\,\lambda_{2}^4\,\lambda_{3}^4+234\,\lambda_{1}\,\lambda_{2}^3\,\lambda_{3}^4+428\,\lambda_{1}^2\,\lambda_{2}^2\,\lambda_{3}^4+306\,\lambda_{1}^3\,\lambda_{2}\,\lambda_{3}^4+75\,\lambda_{1}^4\,\lambda_{3}^4 \\
			&  \quad+16\,\lambda_{2}^5\,\lambda_{3}^3+234\,\lambda_{1}\,\lambda_{2}^4\,\lambda_{3}^3+900\,\lambda_{1}^2\,\lambda_{2}^3\,\lambda_{3}^3+1374\,\lambda_{1}^3\,\lambda_{2}^2\,\lambda_{3}^3+885\,\lambda_{1}^4\,\lambda_{2}\,\lambda_{3}^3+196\,\lambda_{1}^5\,\lambda_{3}^3+42\,\lambda_{1}\,\lambda_{2}^5\,\lambda_{3}^2+428\,\lambda_{1}^2\,\lambda_{2}^4\,\lambda_{3}^2 \\
			&  \quad+1374\,\lambda_{1}^3\,\lambda_{2}^3\,\lambda_{3}^2+1861\,\lambda_{1}^4\,\lambda_{2}^2\,\lambda_{3}^2+1082\,\lambda_{1}^5\,\lambda_{2}\,\lambda_{3}^2+210\,\lambda_{1}^6\,\lambda_{3}^2+34\,\lambda_{1}^2\,\lambda_{2}^5\,\lambda_{3}+306\,\lambda_{1}^3\,\lambda_{2}^4\,\lambda_{3}+885\,\lambda_{1}^4\,\lambda_{2}^3\,\lambda_{3} \\
			&  \quad+1082\,\lambda_{1}^5\,\lambda_{2}^2\,\lambda_{3}+556\,\lambda_{1}^6\,\lambda_{2}\,\lambda_{3}+87\,\lambda_{1}^7\,\lambda_{3}+9\,\lambda_{1}^3\,\lambda_{2}^5+75\,\lambda_{1}^4\,\lambda_{2}^4+196\,\lambda_{1}^5\,\lambda_{2}^3+210\,\lambda_{1}^6\,\lambda_{2}^2+87\,\lambda_{1}^7\,\lambda_{2}+7\,\lambda_{1}^8, \\
			{\eta _3} &= 10\,\lambda_{2}^2\,\lambda_{3}^5+11\,\lambda_{1}\,\lambda_{2}\,\lambda_{3}^5+3\,\lambda_{1}^2\,\lambda_{3}^5+62\,\lambda_{2}^3\,\lambda_{3}^4+200\,\lambda_{1}\,\lambda_{2}^2\,\lambda_{3}^4+182\,\lambda_{1}^2\,\lambda_{2}\,\lambda_{3}^4+51\,\lambda_{1}^3\,\lambda_{3}^4+62\,\lambda_{2}^4\,\lambda_{3}^3+450\,\lambda_{1}\,\lambda_{2}^3\,\lambda_{3}^3 \\
			&  \quad+972\,\lambda_{1}^2\,\lambda_{2}^2\,\lambda_{3}^3+780\,\lambda_{1}^3\,\lambda_{2}\,\lambda_{3}^3+204\,\lambda_{1}^4\,\lambda_{3}^3+10\,\lambda_{2}^5\,\lambda_{3}^2+200\,\lambda_{1}\,\lambda_{2}^4 \,\lambda_{3}^2+972\,\lambda_{1}^2\,\lambda_{2}^3\,\lambda_{3}^2+1762\,\lambda_{1}^3\,\lambda_{2}^2\,\lambda_{3}^2 \\
			&  \quad+1277\,\lambda_{1}^4\,\lambda_{2}\,\lambda_{3}^2+300\,\lambda_{1}^5\,\lambda_{3}^2+11\,\lambda_{1}\,\lambda_{2}^5\,\lambda_{3}+182\,\lambda_{1}^2\,\lambda_{2}^4\,\lambda_{3}+780\,\lambda_{1}^3\,\lambda_{2}^3\,\lambda_{3}+1277\,\lambda_{1}^4\,\lambda_{2}^2\,\lambda_{3}+833\,\lambda_{1}^5\,\lambda_{2}\,\lambda_{3} \\
			&  \quad+165\,\lambda_{1}^6\,\lambda_{3}+3\,\lambda_{1}^2\,\lambda_{2}^5+51\,\lambda_{1}^3\,\lambda_{2}^4+204\,\lambda_{1}^4 \,\lambda_{2}^3+300\,\lambda_{1}^5\,\lambda_{2}^2+165\,\lambda_{1}^6\,\lambda_{2}+21\,\lambda_{1}^7, \\
			{\eta _4} &= 2\,\lambda_{2}\,\lambda_{3}^5+38\,\lambda_{2}^2\,\lambda_{3}^4+51\,\lambda_{1}\,\lambda_{2} \,\lambda_{3}^4+13\,\lambda_{1}^2\,\lambda_{3}^4+88\,\lambda_{2}^3\,\lambda_{3}^3+339\,\lambda_{1}\, \lambda_{2}^2\,\lambda_{3}^3+355\,\lambda_{1}^2\,\lambda_{2}\,\lambda_{3}^3+106\,\lambda_{1}^3\,\lambda_{3}^3+ 38\,\lambda_{2}^4\,\lambda_{3}^2 \\
			&  \quad+339\,\lambda_{1}\,\lambda_{2}^3\,\lambda_{3}^2+890\,\lambda_{1}^2\, \lambda_{2}^2\,\lambda_{3}^2+836\,\lambda_{1}^3\,\lambda_{2}\,\lambda_{3}^2+240\,\lambda_{1}^4\,\lambda_{3}^2+ 2\,\lambda_{2}^5\,\lambda_{3}+51\,\lambda_{1}\,\lambda_{2}^4\,\lambda_{3}+355\,\lambda_{1}^2\,\lambda_{2}^3\, \lambda_{3}+836\,\lambda_{1}^3\,\lambda_{2}^2\,\lambda_{3} \\
			&  \quad+725\,\lambda_{1}^4\,\lambda_{2}\,\lambda_{3}+185\, \lambda_{1}^5\,\lambda_{3}+13\,\lambda_{1}^2\,\lambda_{2}^4+106\,\lambda_{1}^3\,\lambda_{2}^3+240\,\lambda_{1} ^4\,\lambda_{2}^2+185\,\lambda_{1}^5\,\lambda_{2}+35\,\lambda_{1}^6, \\
			{\eta _5} &= 8\,\lambda_{2}\,\lambda_{3}^4+54\,\lambda_{2}^2\,\lambda_{3}^3+87\,\lambda_{1}\,\lambda_{2} \,\lambda_{3}^3+22\,\lambda_{1}^2\,\lambda_{3}^3+54\,\lambda_{2}^3\,\lambda_{3}^2+246\,\lambda_{1}\, \lambda_{2}^2\,\lambda_{3}^2+307\,\lambda_{1}^2\,\lambda_{2}\,\lambda_{3}^2+102\,\lambda_{1}^3\,\lambda_{3}^2+ 8\,\lambda_{2}^4\,\lambda_{3} \\
			&  \quad+87\,\lambda_{1}\,\lambda_{2}^3\,\lambda_{3}+307\,\lambda_{1}^2\,\lambda_{2}^2\, \lambda_{3}+369\,\lambda_{1}^3\,\lambda_{2}\,\lambda_{3}+123\,\lambda_{1}^4\,\lambda_{3}+22\,\lambda_{1}^2\, \lambda_{2}^3+102\,\lambda_{1}^3\,\lambda_{2}^2+123\,\lambda_{1}^4\,\lambda_{2}+35\,\lambda_{1}^5, \\
			{\eta _6} &= 12\lambda_{2}\,\lambda_{3}^3{+}34\lambda_{2}^2\,\lambda_{3}^2{+}65\lambda_{1}\,\lambda_{2} \,\lambda_{3}^2{+}18\lambda_{1}^2\,\lambda_{3}^2{+}12\lambda_{2}^3\,\lambda_{3}{+}65\lambda_{1}\,\lambda_{2}^2 \,\lambda_{3}{+}107\lambda_{1}^2\,\lambda_{2}\,\lambda_{3}{+}45\lambda_{1}^3\,\lambda_{3}{+}18\lambda_{1}^2\, \lambda_{2}^2{+}45\lambda_{1}^3\,\lambda_{2}{+}21\lambda_{1}^4, \\
			{\eta _7} &=8\,\lambda_{2}\,\lambda_{3}^2+8\,\lambda_{2}^2\,\lambda_{3}+18\,\lambda_{1}\,\lambda_{2}\, \lambda_{3}+7\,\lambda_{1}^2\,\lambda_{3}+7\,\lambda_{1}^2\,\lambda_{2}+7\,\lambda_{1}^3, \\
			{\eta _8} &=2\,\lambda_{2}\,\lambda_{3}+\lambda_{1}^2, \\
			{\gamma _0} &=2\,\lambda_{1}\,\left(\lambda_{2}+\lambda_{1}\right)\,\left(\lambda_{3}+\lambda_{1}\right)\, \lambda^2\,\left(\lambda_{2}\,\lambda_{3}+\lambda_{1}\,\lambda_{3}+ \lambda_{1}\,\lambda_{2}\right), \\
			{\gamma _1} &=\left(\lambda_{2}+\lambda_{1}\right)\,\left(\lambda_{3}+\lambda_{1}\right)\,\lambda\,\left(4\,\lambda_{2}\,\lambda_{3}^2+5\,\lambda_{1}\,\lambda_{3}^2+4\, \lambda_{2}^2\,\lambda_{3}+20\,\lambda_{1}\,\lambda_{2}\,\lambda_{3}+12\,\lambda_{1}^2\,\lambda_{3}+5\,\lambda_{1}\, \lambda_{2}^2+12\,\lambda_{1}^2\,\lambda_{2}+\lambda_{1}^3\right), \\
			{\gamma _2}  &= 2\,(2\,\lambda_{2}\,\lambda_{3}^4+2\,\lambda_{1}\,\lambda_{3}^4+10\,\lambda_{2}^2\, \lambda_{3}^3+26\,\lambda_{1}\,\lambda_{2}\,\lambda_{3}^3+15\,\lambda_{1}^2\,\lambda_{3}^3+10\,\lambda_{2}^3\, \lambda_{3}^2+54\,\lambda_{1}\,\lambda_{2}^2\,\lambda_{3}^2+76\,\lambda_{1}^2\,\lambda_{2}\,\lambda_{3}^2+30\, \lambda_{1}^3\,\lambda_{3}^2+2\,\lambda_{2}^4\,\lambda_{3} \\
			&  \quad+26\,\lambda_{1}\,\lambda_{2}^3\,\lambda_{3}+76\, \lambda_{1}^2\,\lambda_{2}^2\,\lambda_{3}+73\,\lambda_{1}^3\,\lambda_{2}\,\lambda_{3}+20\,\lambda_{1}^4\,\lambda_{3} +2\,\lambda_{1}\,\lambda_{2}^4+15\,\lambda_{1}^2\,\lambda_{2}^3+30\,\lambda_{1}^3\,\lambda_{2}^2+20\, \lambda_{1}^4\,\lambda_{2}+3\,\lambda_{1}^5), \\
			{\gamma _3} &= \lambda_{3}^4+16\,\lambda_{2}\,\lambda_{3}^3+18\,\lambda_{1}\,\lambda_{3}^3+34\,\lambda_{2}^ 2\,\lambda_{3}^2+99\,\lambda_{1}\,\lambda_{2}\,\lambda_{3}^2+60\,\lambda_{1}^2\,\lambda_{3}^2+16\,\lambda_{2}^ 3\,\lambda_{3}+99\,\lambda_{1}\,\lambda_{2}^2\,\lambda_{3}+151\,\lambda_{1}^2\,\lambda_{2}\,\lambda_{3}+60\, \lambda_{1}^3\,\lambda_{3}+\lambda_{2}^4 \\
			&  \quad+18\,\lambda_{1}\,\lambda_{2}^3+60\,\lambda_{1}^2\,\lambda_{2}^2+60\, \lambda_{1}^3\,\lambda_{2}+15\,\lambda_{1}^4, \\
			{\gamma _4} &= 2\,\lambda_{3}^3+12\,\lambda_{2}\,\lambda_{3}^2+15\,\lambda_{1}\,\lambda_{3}^2+12\, \lambda_{2}^2\,\lambda_{3}+39\,\lambda_{1}\,\lambda_{2}\,\lambda_{3}+25\,\lambda_{1}^2\,\lambda_{3}+2\,\lambda_{2}^3 +15\,\lambda_{1}\,\lambda_{2}^2+25\,\lambda_{1}^2\,\lambda_{2}+10\,\lambda_{1}^3, \\
			{\gamma _5} &= 6\,\lambda_{3}^2+16\,\lambda_{2}\,\lambda_{3}+22\,\lambda_{1}\,\lambda_{3}+6\,\lambda_{2}^2+ 22\,\lambda_{1}\,\lambda_{2}+15\,\lambda_{1}^2, \\
			{\gamma _6} &= 2\,\left(2\,\lambda_{3}+2\,\lambda_{2}+3\,\lambda_{1}\right), \\
			{\gamma _7} &= 1.
		\end{align*}
		\hrulefill
	\end{figure*}
}

% \bibitem{1}
% C. Cho, C. Lin and J. Wang, ``Reliable grouping gaf algorithm using hexagonal virtual cell structure,'' in \textit{Proc. ICST'08}, Tainan, Nov. 2008.

\end{document}